\documentclass[12pt,preprint]{aastex}

\newcommand{\micronn}{\,\hbox{$\mu$m}}

\newcommand\msun{\hbox{\,M$_\odot$}}

\newcommand\lsun{\hbox{\,L$_\odot$}}
\newcommand\kms{ km~s$^{-1}$}

\newcommand\lamvir{$\lambda$~Vir}
\newcommand\lam{$\lambda$}
\def\kms{\ifmmode{\rm km\thinspace s^{-1}}\else km\thinspace s$^{-1}$\fi}

\shorttitle{$\lambda$ Virginis}
\shortauthors{Zhao et al.}

\begin{document}

\title{Physical Orbit for $\lambda$~Virginis and a Test of Stellar Evolution Models}

\author{M.~Zhao\altaffilmark{1},
J.~D.~Monnier\altaffilmark{1},
G.~Torres\altaffilmark{2},
A.~F.~Boden\altaffilmark{3},
A.~Claret\altaffilmark{4},
R.~Millan-Gabet\altaffilmark{3},
E. Pedretti\altaffilmark{1},
J.-P.~Berger\altaffilmark{5},
W.~A.~Traub\altaffilmark{2},
F.~P.~Schloerb\altaffilmark{6},
N.~P.~ Carleton\altaffilmark{2},
P.~ Kern\altaffilmark{5},
M.~G.~Lacasse\altaffilmark{2},
F.~Malbet\altaffilmark{5},
K.~ Perraut\altaffilmark{5}
}

\altaffiltext{1}{mingzhao@umich.edu: University of Michigan Astronomy Department,
941 Dennison Bldg, Ann Arbor, MI 48109-1090, USA}
\altaffiltext{2}{Harvard-Smithsonian Center for Astrophysics, 60 Garden
St, Cambridge, MA, 02138, USA}
\altaffiltext{3}{Michelson Science Center, California Institute of Technology,
770 South Wilson Avenue, Pasadena, CA 91125}
\altaffiltext{4}{Instituto de Astrof{\'\i}sica de Andaluc{\'\i}a, CSIC, Apartado 3004, E-18080 Granada, Spain}
\altaffiltext{5}{Laboratoire d'Astrophysique de Grenoble, 414 Rue de la Piscine 38400 Saint Martin d'Heres,
France}
\altaffiltext{6}{University of Massachusetts, Amherst}

\begin{abstract}
$\lambda$ Virginis is a well-known double-lined spectroscopic Am
binary with the interesting property that both stars are very
similar in abundance but one is sharp-lined and the other is
broad-lined. We present combined interferometric and spectroscopic
studies of \lamvir. The small scale of the \lamvir\ orbit ($\sim20$ mas) is
well resolved by the Infrared Optical Telescope Array (IOTA),
allowing us to determine its elements as well as the physical properties of the
components to high accuracy. The masses of the two stars are
determined to be
1.897 $\msun$ and 1.721 $\msun$, with 0.7$\%$ and 1.5$\%$
errors respectively, and the two stars are found to have the same
temperature of 8280 $\pm$ 200 K. The accurately determined
properties of \lamvir\ allow comparisons between observations and
current stellar evolution models, and reasonable matches are
found. The best-fit stellar model gives \lamvir\  a subsolar
metallicity of Z=0.0097, and an age of 935 Myr. The orbital and
physical parameters of \lamvir\ also allow us to study its tidal evolution
time scales and status. Although currently atomic
diffusion is considered to be the most plausible cause of the Am
phenomenon, the issue is still being
actively debated in the literature. 
With the present study of the
properties and evolutionary status of \lamvir, this system is an
ideal candidate for further detailed abundance analyses that might
shed more light on the source of the chemical anomalies in these A
stars.

\end{abstract}

\keywords{binaries:spectroscopic --- binaries:visual --- stars:fundamental
parameters --- stars: individual ($\lambda$ Virginis) --- instrumentation:
interferometers }

\section{Introduction}
Am stars were first recognized by \citet{Titus1940} as a group of
stars for which spectral classification is
ambiguous. The Ca II K lines correspond to earlier types than derived
from the Balmer lines, which in turn give earlier types than the
metallic lines.
Am stars
generally have deficient CNO abundances \citep[e.g.,][
etc.]{Roby1990, Sadakane1989}, while their iron peak and rare
earth elements are generally overabundant \citep{Veer1988, Cayrel1991}.
Statistical studies \citep{Abt1961, Abt1995, Abt2000} suggest that
virtually all Am stars are binaries with  projected equatorial
rotational velocities less than 120~\kms, and
it is the slow rotation that causes the abundance anomalies of Am stars.
It is now widely believed that atomic diffusion in slowly
rotating stars (e.g., Am and Ap stars) will occur in an outer
convection zone so that some elements will be depleted in the
atmosphere while others will become overabundant, which partly
explains the chemical peculiarity of these
stars \citep{Michaud1980, Richer1998}. Recent progress has been
made on atomic diffusion models \citep{Richer2000}, and
\citet{Michaud2005} have shown an example study of $o$~Leo
indicating that these models can produce
abundance anomalies that are consistent with observations.
 However, the masses they adopted from \citet{Griffin2002}
have much larger error bars (more than 20 times larger) than the
original determinations of \citet{Hummel2001} and no explanation
was given on such a large difference. This implies that if the
values from \citet{Griffin2002} were wrong, the studies of
\citet{Michaud2005} would be affected and their conclusions might
be changed as well. Very recently, \citet{Bohm-Vitense:06} 
studied the interaction between Am stars and the interstellar
medium, and suggested that the Am phenomenon may be due at least in
part to accretion of interstellar material rather than the more
popular explanation in terms of atomic diffusion processes.
This study challenges the most popular explanation of the Am
phenomenon and makes this puzzle more interesting yet still
unclear. Although Am stars have been studied intensively since
their discovery, only a few of them have well determined
properties. Therefore, in order to address these problems, more
precise and accurate measurements of Am stars are 
required so that more detailed studies can be conducted to help
improve our understanding of the role of atomic diffusion and,
eventually, the cause of the abundance anomalies in Am stars.

$\lambda$ Virginis (HD 125337, HIP 69974, HR 5359; $V=4.523$ mag, $H=4.282$ mag) was
first reported to be a double-lined spectroscopic binary by
\citet{Moore1911}. The two components were classified as
metallic-lined A (Am) stars \citep{cowley1969, Levato1975}.
 Early spectroscopic studies estimated its
orbital parameters and found a period of 206 days with very low
eccentricity ($\sim0.079$) \citep{Colacevich1941, Abt1961,
Stickland1975, Stickland1990}. Chemical abundance studies
\citep{Colacevich1941, Stickland1975} suggested the interesting
property of \lamvir\ that both stars are very similar in abundance
despite their different rotation velocities with the primary being
broad-lined (with $v \sin i = 35$~\kms) and the secondary sharp-lined
(with $v \sin i = 16$~\kms). The differing rotation rates and the
unusual metallic-lined nature of the system, as well as the
similarity in the abundance of the two components give us a unique
opportunity to test stellar models and study its evolutionary status.

In this paper, we report the combined interferometric and
spectroscopic study of \lamvir\ and the testing of stellar
evolution models. The observations span several orbital periods,
providing enough orbital coverage and allowing us to deduce the
orbital and physical properties of the system precisely. After
describing the observations in \S\ref{obs}, we present the orbit
determination in \S\ref{orb_determine}, including the discussion of bandwidth
smearing effect for the interferometric visibilities and biases in closure phase
measurements. We determine its physical properties in \S\ref{sec:physics},
and compare the resulting properties with stellar models in \S\ref{sec:models} and tidal
evolution theory in \S\ref{sec:tidal}. Finally, we
give our conclusions and summary in \S\ref{sec:summary}.

\section{Observations}
\label{obs}

\subsection{Spectroscopic observations and reductions}
\label{sec:spectroscopy}
The spectroscopic observations of \lamvir\ were conducted at the
Harvard-Smithsonian Center for Astrophysics (CfA) between 1982 July
and 1991 February, mostly with an echelle spectrograph on the 1.5-m
Wyeth reflector at the Oak Ridge Observatory (Harvard, Massachusetts).
A single echelle order was recorded with an intensified Reticon diode
array giving a spectral coverage of about 45\,\AA\ at a central
wavelength of 5188.5\,\AA. The main spectral feature in this region is
the \ion{Mg}{1}~b triplet, although there are numerous other metallic
lines as well. The resolving power is $\lambda/\Delta\lambda\approx
35,\!000$. Occasional observations were made also with nearly
identical instruments on the 1.5-m Tillinghast reflector at the F.\
L.\ Whipple Observatory (Mt.\ Hopkins, Arizona) and the Multiple
Mirror Telescope (also on Mt.\ Hopkins, Arizona), prior to its
conversion to a monolithic mirror. A total of 130 spectra were
collected, with signal-to-noise ratios (SNRs) ranging from 20 to about
50 per resolution element of 8.5~\kms.

Radial velocities were derived using TODCOR \citep{Zucker:94}, a
two-dimension\-al cross-correlation algorithm well suited to our
relatively low SNR spectra. TODCOR uses two templates, one for each
component of the binary, and significantly reduces systematics due to
line blending that are often unavoidable in standard one-dimensional
cross-correlation techniques \citep[see, e.g.,][]{Latham:96}.  The
templates were selected from a large library of synthetic spectra
based on model atmospheres by R.\ L.\ Kurucz (available at
\url{http://cfaku5.cfa.harvard.edu}), computed for us by Jon Morse
\citep[see also][]{Nordstrom:94,Latham:02}. These calculated spectra
are available for a wide range of effective temperatures ($T_{\rm
eff}$), projected rotational velocities ($v \sin i$), surface
gravities ($\log g$) and metallicities. Experience has shown that
radial velocities are largely insensitive to the surface gravity and
metallicity adopted for the templates. Consequently, the optimum
template for each star was determined from grids of cross-correlations
over broad ranges in temperature and rotational velocity, seeking to
maximize the average correlation weighted by the strength of each
exposure \citep[see][]{Torres:02}. For the surface gravity we adopted
the value of $\log g = 4.0$ for both stars (see \S\ref{sec:models}),
and for the metallicity we initially adopted the solar
composition. However, in view of the metallic-lined nature of the
stars we repeated the procedure for a range of metallicities from
[m/H] $= -1.0$ to [m/H] $= +0.5$ in steps of 0.5 dex. We found the
best match to the observed spectra for [m/H] $= +0.5$, which is
consistent with the enhanced surface abundances expected for these
objects. At this metallicity the effective temperatures we derive are
$8800 \pm 200$~K for both stars, and the rotational velocities are
$v_1 \sin i = 36 \pm 1$~\kms\ and $v_2 \sin i = 10 \pm 2$~\kms\ for
the primary and secondary, respectively. The rotational velocity
estimates are fairly consistent with determinations by other authors:
\cite{Stickland1975} reported 35~\kms\ and 16~\kms\ (no uncertainties
given), and \cite{Abt1995} estimated 31~\kms\ and 13~\kms, with
uncertainties of about 8~\kms. Very rough values without uncertainties
were estimated more recently by \cite{Shorlin:02} as $\sim$50~\kms\
and $< 10$~\kms.  We discuss the temperature estimates in
\S\ref{sec:physics}.

In addition to the radial velocities and stellar parameters, we
derived the spectroscopic light ratio following \cite{Zucker:94}. The
result, $\ell_2/\ell_1 = 0.58 \pm 0.02$, corresponds to the mean
wavelength of our observations (5188.5\,\AA) and is not far from the
visual band.

Due to the narrow wavelength coverage of the CfA spectra there is
always the possibility of systematic errors in the velocities,
resulting from lines of the stars moving in and out of the spectral
window with orbital phase \citep{Latham:96}.  Occasionally these
errors are significant, and experience has shown that this must be
checked on a case-by-case basis \citep[see, e.g.,][]{Torres:97,
Torres:00}. For this we performed numerical simulations in which we
generated artificial composite spectra by adding together synthetic
spectra for the two components, with Doppler shifts appropriate for
each actual time of observation, computed from a preliminary orbital
solution.  The light ratio adopted was that derived above.  We then
processed these simulated spectra with TODCOR in the same manner as
the real spectra, and compared the input and output velocities.
Although the differences for \lamvir\ were well under 1~\kms, they are
systematic in nature and we therefore applied them as corrections to
the raw velocities for completeness.  The final velocities including
these corrections are given in Table~\ref{tab:rvs}.  Similar
corrections were derived for the light ratio, and are already
accounted for in the value reported above.

The stability of the zero-point of the velocity system was monitored
by means of exposures of the dusk and dawn sky, and small run-to-run
corrections were applied in the manner described by \cite{Latham:92}.
These corrections are also included in Table~\ref{tab:rvs}. The
accuracy of the CfA velocity system, which is within about 0.14~\kms\
of the reference frame defined by minor planets in the solar system,
is documented in the previous citation and also by \cite{Stefanik:99}
and \cite{Latham:02}.

\subsection{Interferometric Observations and Data Reduction}
\label{obs2} The interferometric observations of \lamvir\  were
carried out using the Infrared Optical Telescope Array
\citep[IOTA,][]{Traub2003} also at the F. L. Whipple Observatory. 
IOTA is a three
0.45m-telescope interferometer array that is movable along its
L-shaped southeast and northeast arms, providing several different
array configurations and having baselines up to 38m. Light from
each telescope is focused into a single-mode fiber and the beams
from 3 fibers are split and combined by the ``pair-wise'' beam
combiner IONIC-3 \citep{Berger2003SPIE} to form six fringes.
Fringes are temporally scanned by piezo scanners in the delay
lines, and are then detected by a PICNIC camera
\citep{Pedretti2004}. This detection scheme leads to high
sensitivities of IOTA \citep[$\sim$7th magnitude at $H$ band,][]{monnier2004}
and allows for precise measurements of visibilities and
closure phases.

The observations reported here were taken in the $H$ band (\lam$_0$
= 1.647\micronn, $\Delta$\lam = 0.30\micronn) between 2003
February and 2005 June, spanning four orbital periods (853 days)
and covering a broad range of orbital phases, and different array
configurations were applied to obtain good $uv$ coverage. The
observations were carried out following
 the standard procedures
\citep[e.g.,][]{monnier2004}, and the observation log is
listed in Table~\ref{table_obslog}. In short, \lamvir\  was
observed in conjunction with nearby unresolved calibrators (HD
126035, HD 129502, HD 158352) to calibrate the varying system
visibilities and closure phases caused by the instrumental
response and the effect of atmospheric seeing. Each single
observation typically consists of 200 scans within $\sim$4 min,
followed by calibration measurements of the background and
individual response of each telescope. Two different piezo scan
modes were used for different observing runs (see
Table~\ref{table_obslog}), mode one before 2003 June 17th
(Telescope A fixed, Telescope B scan range: 50.8 $\micron$,
Telescope C scan range: 25.4 $\micron$), and mode two thereafter
(Telescope A fixed, Telescope B scan range: 25.4 $\micron$,
Telescope C scan range: $-$25.4 $\micron$). The effect of different
scanning modes is discussed later in the Appendix.

Reduction of the
squared-visibilities ($V^2$) and the closure phases was carried out using
established IDL routines described by
\citet{monnier2004, monnier2006}.
 In short,
we measure the power spectrum of each interferogram which is
proportional to the broad-band {$V^2$} 
\citep[see][for an outline of the method]{CRM1997}, and correct for intensity
fluctuations as well as bias terms that stem from read noise,
background noise, etc. The variable flux ratios of each baseline
are calibrated using a flux transfer matrix \citep{monnier2006}.
Measurement errors are obtained from the scatter of the data and
are then combined with calibration errors. The calibration error,
established statistically from the data fitting procedures (see
\S4), is $\sim2\%$ for {$V^2$}, corresponding to $1\%$ error
in the visibility. In order to measure the closure phases, a
real-time fringe-tracking algorithm \citep{Pedretti2005} was
applied to ensure that the interferograms are detected
simultaneously in nearly all baselines (at least two are detected
if fringes in the third baseline is weak). The closure phases are
then obtained by calculating and averaging the bi-spectrum (triple
product) in complex space, with the frequencies of each triple
product closed, i.e., $\nu_{AB}+\nu_{BC}+ \nu_{CA}=0$
\citep{Baldwin1996}. The instrumental closure phase offset
\citep[$\leq 0.5{\arcdeg}$,][]{monnier2006} is calibrated by using
unresolved calibrators listed in the observation log. The
calibration errors of the closure phases are dominated by fluctuations
that result from extra optical path differences (OPDs) caused by
the atmospheric piston fluctuations. We will discuss this effect
in the Appendix and the error estimation for the closure phases in
\S\ref{cp}.

\section{Orbit Determination}
\label{orb_determine}

\subsection{Bandwidth Smearing Effect of $V^2$}
\label{vis}
Interferometric measurements use a finite range of bandwidth. The
resulting fringe packets thus suffer a modulation
 in the amplitude due to the overlap of
fringes with different wavelengths, especially at the edges of the
packets. For binary stars, the observed interferogram results from
the interference of two fringe packets with an interferometric
delay of $\vec{B}\cdot\vec{\rho}$ due to the binary separation
(where $\vec{B}$ is the projected baseline vector $(B_x,B_y)$ in meters
and
 $\vec{\rho}$ is the angular separation ($a$, $b$) of the binary in units of
radians).  Because the two fringe packets are modulated by
bandwidth smearing, the resulting observed interferogram is also
affected by this, causing significant systematic errors to
the measured visibilities and closure phases. This effect is
pronounced for broad band filters such as the $H$-band filter of the
IOTA PICNIC camera. Our preliminary binary modelling indicated a
poor fit to the squared-visibilities and the closure phases,
evidenced by a large reduced $\chi^2$ ($\chi^2_{\nu}$). Therefore,
before we determined the orbit of \lamvir, we first investigated
the influence of bandwidth smearing on our data.

\begin{figure}[thb]
\begin{center}
{
\includegraphics[angle=0,width=3.2in]{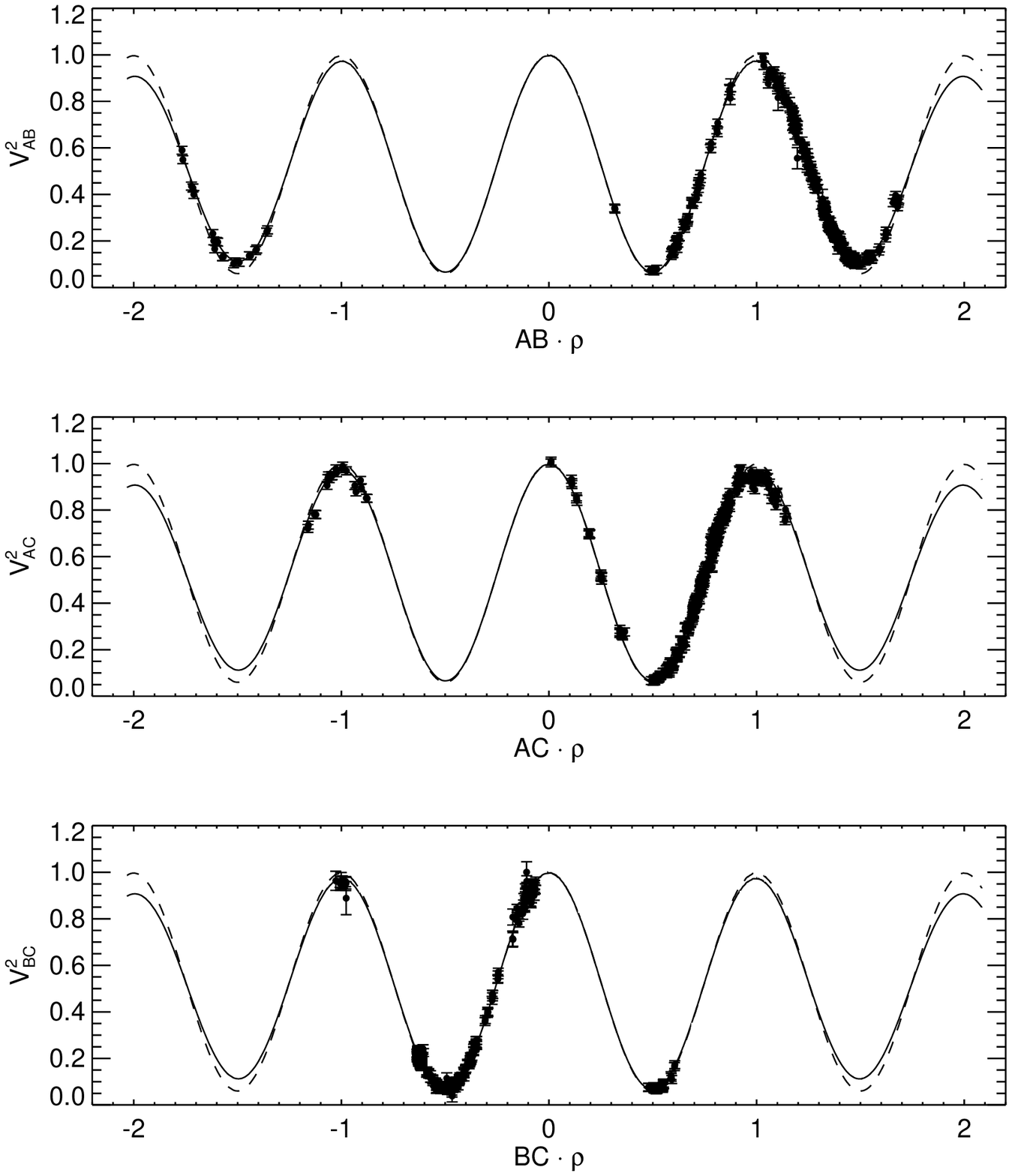}
\includegraphics[angle=0,width=3.2in]{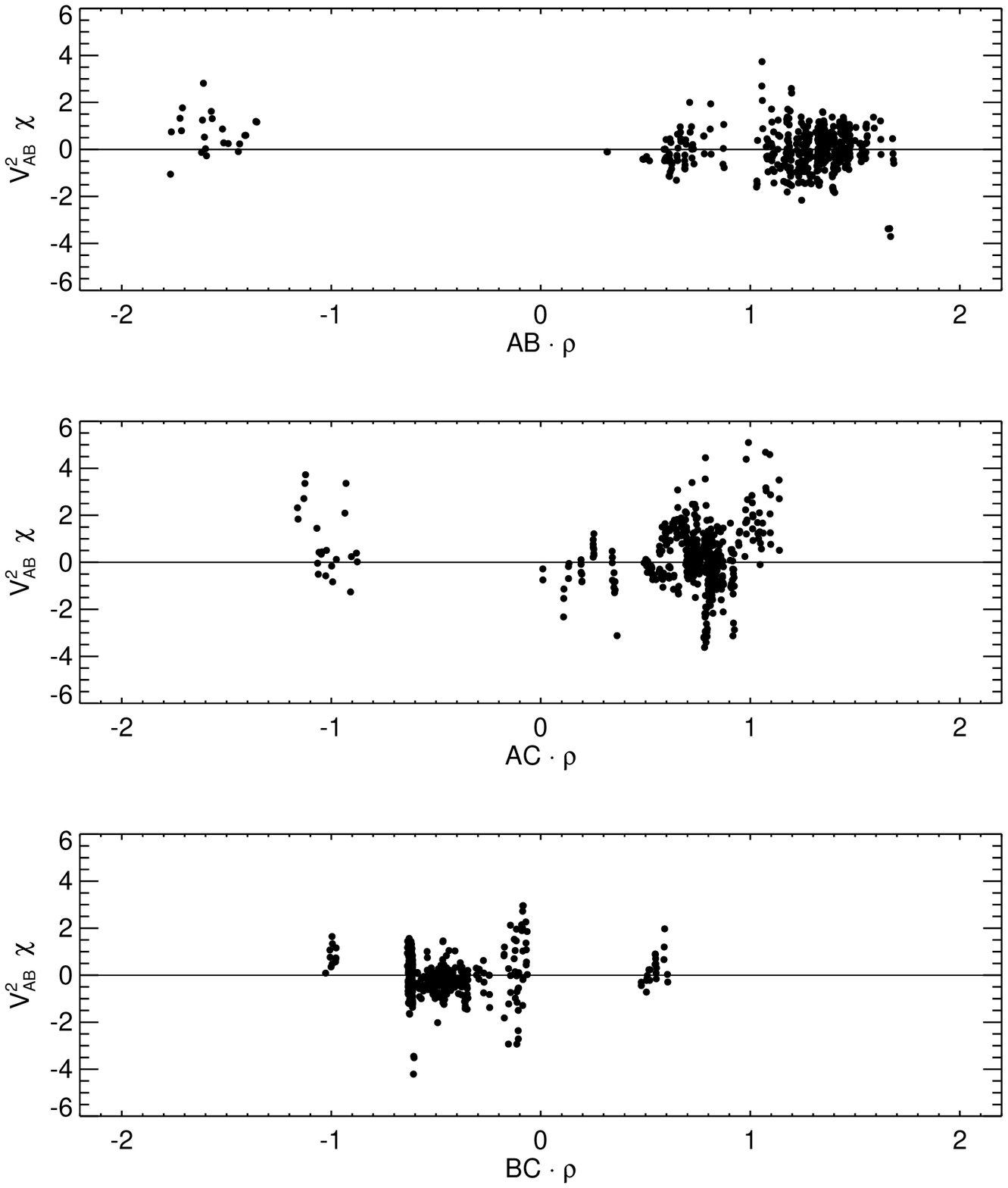}}
\hphantom{.....}
\caption{ 
$V^2$ for 3 IOTA baselines vs. interferometric delay ($\vec{B}
\cdot\vec{\rho} $) in units of wavelength. The dashed lines indicate the
original squared-visibility model with no bandwidth smearing
correction, while the solid lines show the models corrected for
bandwidth smearing. $V^2$ data are overplotted with error-bars of
1-$\sigma$. The corresponding normalized residuals for the corrected model
(i.e., residual/error) are shown in the right panels for the 3 baselines
respectively. \label{bandwidth2}}
\end{center}
\end{figure}

The standard
monochromatic squared-visibility of a binary can be written as
\begin{equation}
V^2  =  \frac{ {|V_1|}^2 + r^2{|V_2|}^2 + 2r \cdot |V_1|\cdot|V_2|\cdot
\cos{\frac{2\pi}{\lambda} \vec{B} \cdot \vec{\rho} }}{{(1 + r)}^2},\label{eq1}
\end{equation}
where $r$ is the flux ratio, and $V_1$, $V_2$ are the visibilities
of the primary and the secondary respectively \citep{summer1999}.
For the case of IOTA IONIC-3, where we measure the power-spectrum of the interferogram
 to determine the broad-band squared-visibility\footnote{This is equivalent
to integrating the squared-visibility over the full wavelength range to get the broad-band
value.}\citep[see e.g.,][]{CRM1997}, 
we integrate the squared-visibility over
the whole bandpass and subtract Equation \ref{eq1} from it
to obtain the difference between the polychromatic and the
monochromatic squared-visibilities:
\begin{equation}
\Delta V^2 = V_{BS}^2 - V^2
=\frac{2r \cdot |V_1| \cdot |V_2| \cdot \cos(2 \pi
\delta) \cdot (\exp{\frac{-\delta^2}{2 {f}^2}} - 1)}{(1 +r)^2
},
\label{eq2}
\end{equation}
where \begin{equation}
f = \frac{\lambda \cdot \beta}{\Delta \lambda \cdot \sqrt{8 \ln2}}.
\label{eq4}
\end{equation}
We used a gaussian envelope function, $\exp{\frac{-\delta^2}{2 {f}^2}}$,
 to approximate the modulation of the interferogram, where
$\delta =\frac{\vec{B} \cdot \vec{\rho}}{\lambda}= \frac{B_xa + B_yb}{\lambda}$
 is the phase difference of the two components
in unit of wavelength, $\beta$ is the introduced bandpass coefficient, and
$f$ is the corresponding bandwidth smearing coefficient which is
also 1 $\sigma$ of the envelope function of the interferogram. The
exact value of $f$ depends on the bandpass shape and windowing
function. For example, for a ``top-hat'' bandpass approximation,
$f \simeq $4.0; for a Gaussian bandpass approximation with
FWHM=$\Delta \lambda$, $f \simeq $ 2.6. We applied Equation
\ref{eq2} to our squared-visibility model with $f$ being a free
parameter. The new best-fit is significantly improved (
$\chi^2_{\nu} \sim 1.3$) compared to the preliminary result
($\chi^2_{\nu} \sim 1.9$), giving $f$ a value of 3.4 which is
consistent with the fact that the bandpass of IOTA is between a
``top-hat'' and a Gaussian function. Fig.~\ref{bandwidth2} shows
the best-fit squared-visibility models before and after applying
the bandwidth smearing correction. The data are plotted versus
interferometric delay $\vec{B}\cdot\vec{\rho}$ (i.e., projected
baseline $\times$ angular separation of the binary) in unit of wavelength. The
corresponding normalized residuals (i.e., normalized by their errors) are shown in the
left panels for the 3 baselines respectively. As can be seen, the
visibility amplitudes around  $\pm1.0$ and $\pm1.5$ wavelengths in
baseline AB (the top panel) are reduced a large amount from the
original sinusoidal $V^2$ model because of the bandwidth smearing
effect and the applied correction improved the fit significantly.
Baselines AC and BC are shorter than baseline AB, and therefore
provide measurements with delay differences $<1$ wavelength
 and suffer less amplitude reduction than baseline AB.

A group of data around 1 wavelength in baseline AC from two different
observations (2003Feb17 and 2004Apr) have large normalized residuals ($>5$)
even after removing all known calibration errors. The orbital phases of
these two epochs are $\sim$0.1 and $\sim$0.20--0.25, respectively.
Inspection of these data revealed unusually high variations
in the system visibilities on this baseline, indicating the poor fit at these
epochs is likely due to calibration problems rather than errors in our determined
orbital parameters.

\subsection{Bandwidth Smearing Effect of Closure Phase and OPD Fluctuations}
\label{cp}
Our preliminary best-fit on closure phases also showed large residuals,
leading to even larger
$\chi^2_{\nu}$ ($\sim3$) than that of the squared-visibilities. This can also be the
result of the bias induced by bandwidth smearing.
However, unlike the case for the visibilities, this bias in the closure phases
does not have a particularly simple analytical expression.
One can only simulate this bias numerically, making it more difficult to
look into the influence of bandwidth smearing. In our approach, we simulate the
observational data of \lamvir\ by generating 3 interferograms
for the 3 IOTA baselines at different epochs. The different piezo scan modes are
also taken into account. We then put the 3 interferograms into the
IOTA data reduction pipeline (\S\ref{obs2}) to reproduce the ``measured''
closure phases as in
real observations. We adopted the same bandpass function and
bandwidth smearing coefficient from the visibility modelling
(\S\ref{vis}). By varying the width of the interferogram
envelope function, we simulated the closure phases for both the monochromatic
and the polychromatic cases.

Fig.~\ref{smearing1} shows the bandwidth-smearing corrected (dotted line) and
the original un-corrected (solid line) closure phase models for two typical
observations (2003Mar24 $\&$ 2005Jun16). These two observations represent
two different a scanning modes, mode 1 for 2003Mar24 (left panel) and mode 2 for
the latter one (right panel).
Fig.~\ref{smearing1} indicates that bandwidth
smearing can change the closure phases by $\sim$ 5 degrees at these two epochs.
Although the fit is improved in the right panel by the simulated bandwidth
smearing model, the one in the left is worse than the original fit.
In fact, the original model deviates from the measured closure phases by up to
10 degrees in the whole data set, and the simulated
bandwidth smearing cannot reduce these deviations significantly, implying
other biases may exist in the closure phase measurements.

Another source of error in the closure phases stems from the
offsets of the fringe phases due to extra OPDs induced by the
atmospheric piston fluctuations. Further investigations (see the
Appendix) suggest that this effect does dominate the errors of our
closure phase measurements. To reduce the influence of this effect
on our fits, we estimate the errors of the closure phases based on
their uncertainties obtained from the simulations of closure phase
fluctuations caused by extra OPDs. The details of the simulation
and the corresponding closure phase behaviors are discussed in the
Appendix. Fig.~\ref{smearing3} shows the best-fit closure phase
model for the two typical observations, overplotted with the
observed data and the estimated errors. The errors in the first
epoch (left panel) are much smaller than those in the second epoch
due to their differing scan modes. The best fit leads to a
$\chi^2_{\nu}$ of 1.2 with 476 degrees of freedom for the closure phases (previously
$\chi^2_{\nu}\sim3$), which is significantly reduced as a result
of the reliable error estimation.

\subsection{The final orbit}
\label{finalorb}

With the bandwidth smearing effects addressed as described above, we
proceed in this section with a simultaneous Keplerian orbital fit to
the radial velocities, the squared visibilities, and the closure
phases for \lamvir. This allows us to determine the full set of 
orbital elements, for which the closure phases remove the ambiguity in
the position angle of the ascending node ($\Omega$) that is usually   
inherent in the visibility measurements. The inclination angle $i$ is 
determined from the interferometric data, and consequently the masses 
$M_1$ and $M_2$ can be found from the spectroscopic values of $M_1\sin^3i$
and $M_2\sin^3i$. 
Since neither of the \lamvir\  components are resolved by IOTA, we take the sizes of the two components
into account by using a uniform stellar disk model
\citep{summer1999}. The applied diameters, 0.40 mas
 for the primary and 0.30 mas for the secondary, are consistent with the values
determined in \S\ref{sec:models}.
The overall $\chi^2$ of
the measurements is minimized using standard non-linear least-squares
techniques, in our case the Levenberg-Marquardt algorithm, and the 
errors of the best-fit parameters are estimated using the bootstrap
method \citep{press1992}.

\begin{figure}[thb]
\begin{center}
{
\includegraphics[angle=90,width=3.4in]{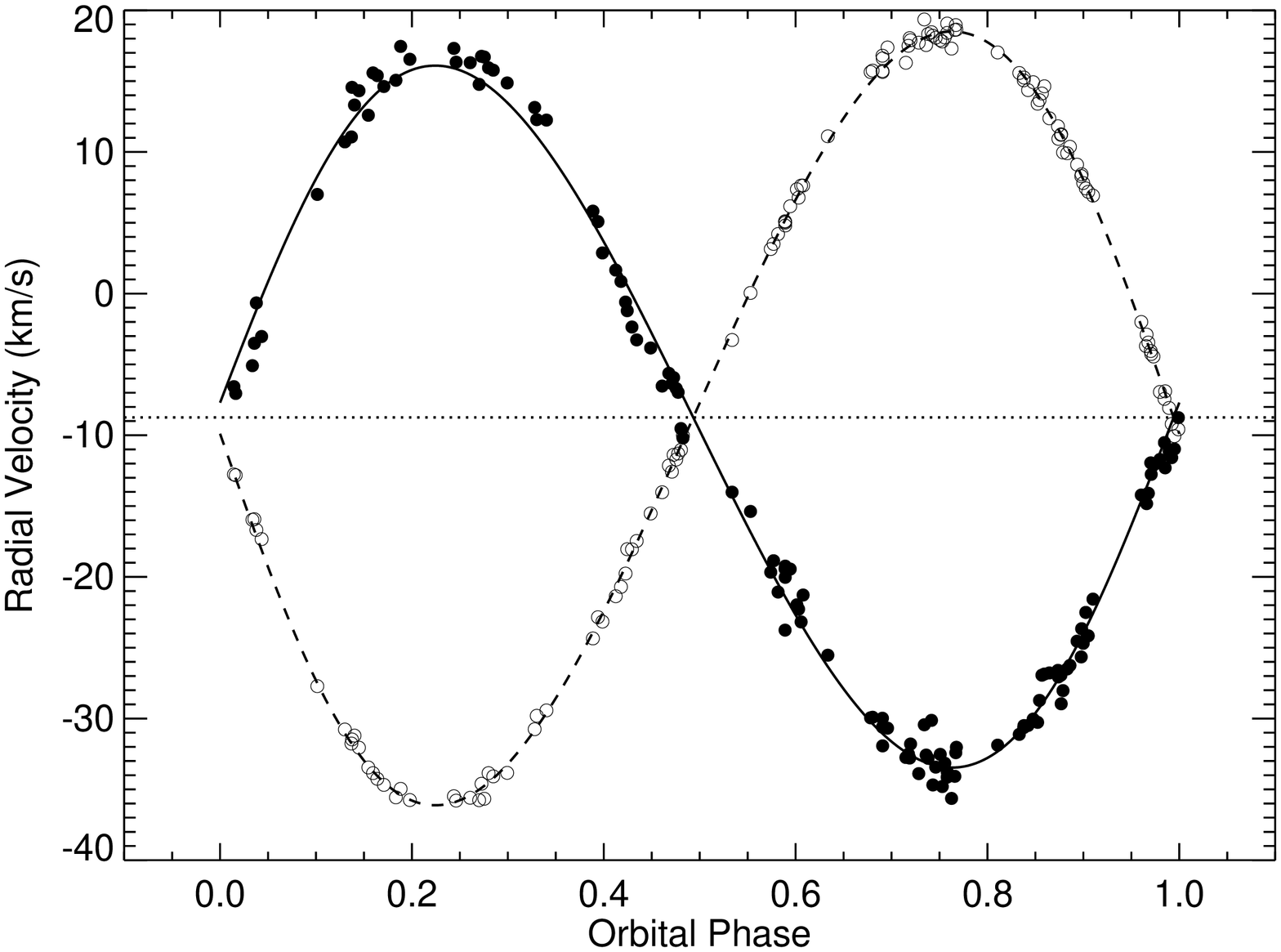}
\includegraphics[angle=90,width=2.8in]{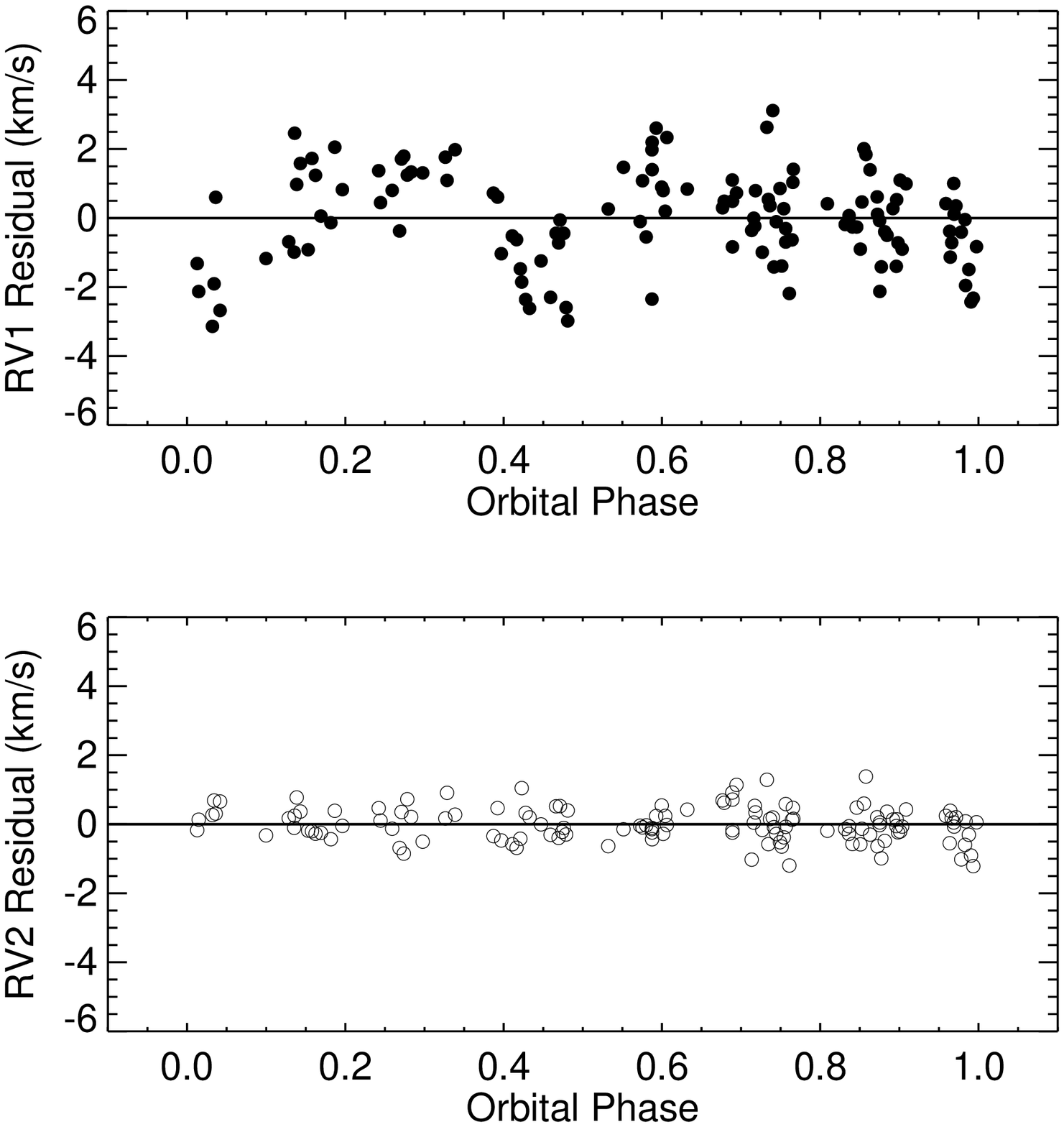}}
\hphantom{.....}

\caption{
Best-fit radial velocity model vs. orbital phase. The data are shown with filled circles for
the primary and open circles for the secondary respectively. The best-fit radial velocity curves are
also shown (primary: solid line, secondary: dashed line). The dotted line
indicates the systemic velocity of the primary. 
Secondary velocities have been
corrected for the offset described in the text.
Velocity residuals
are given in the right panels. The larger values for the primary are
caused by the larger rotational broadening of its spectral lines, 
and possibly also by template mismatch due to the anomalous
abundances (see text).
\label{rvplot}}
\end{center}
\end{figure}

The calibration error of the squared-visibilities is obtained from
the fitting procedures, leading to a value of $\Delta V^2=0.017$ that corresponds to a $1.7\%$
error for an unresolved source ($V^2=1$). Closure phase errors are determined
in the previous section (\S\ref{cp}).
The statistical weights (or errors) of the radial velocity data are established
from the model fitting procedures as well.
In particular, we start with initial values and iterate the $\chi^2$
minimization for the primary and the secondary radial velocities
until the estimated weights converge. The resulting error for the primary,
1.34~\kms, is larger than that of the secondary, 0.50~\kms, due to
the fact that the primary is broad-lined and the secondary is sharp-lined.
 Fig.~\ref{rvplot} shows the radial velocity models, plotted versus orbital phase.
In the fit we allowed for a     
possible offset between the primary and secondary velocities that
could originate from a template mismatch in the cross-correlations due
to the metallic-lined nature of the stars (see \S\ref{sec:spectroscopy}). We found a
small but statistically significant offset of $0.70 \pm 0.13$~\kms,
which has been accounted for in plotting the secondary velocities. 
The corresponding best-fit residuals are given in the
right panels. It is noticeable that the primary has much larger
residuals than the secondary. We searched carefully for the
presence of a third star that might be responsible for fluctuations in the
orbit. However, neither the spectroscopic or interferometric
data,  nor the {\it Hipparcos\/} data and other available online catalogs
(such as 2MASS) indicate any such evidence. Keck aperture masking
was also used on this source and no wider companions ($\rho <
0.5''$) were observed at $2\mu m$ (Monnier 2005, private
communication). These investigations indicate the absence of a
third companion within the detection limits, and even if it exists,
it would have negligible influence on the \lamvir\ orbit.
The temperature and luminosity of the primary star are typical of
$\delta$~Sct variables, which have pulsation periods usually less than
0.3 days. Our velocity sampling is not well suited to discover
periodicities as short as this. However, it is unlikely that
oscillations of this kind contribute significantly to the velocity
residuals we see in Fig.\ 4.  Instead, the pattern suggests a much
longer-period variation (quite apparent in the figure, at least
between phase 0.0 and 0.5). Indeed, a periodogram analysis of the
residuals shows significant power at a period very close to half the
orbital period. We believe the source of these residuals is template
mismatch, caused by the anomalous abundances of the stars. The primary is more 
vulnerable to these effects due to its broader lines. The
dependence with phase comes from the unavoidable fact that different
spectral lines shift in and out of our spectral window as the stars
orbit each other.

\begin{figure}[thb]
\begin{center}
{
\includegraphics[angle=0,width=5.35in, height=5.6in]{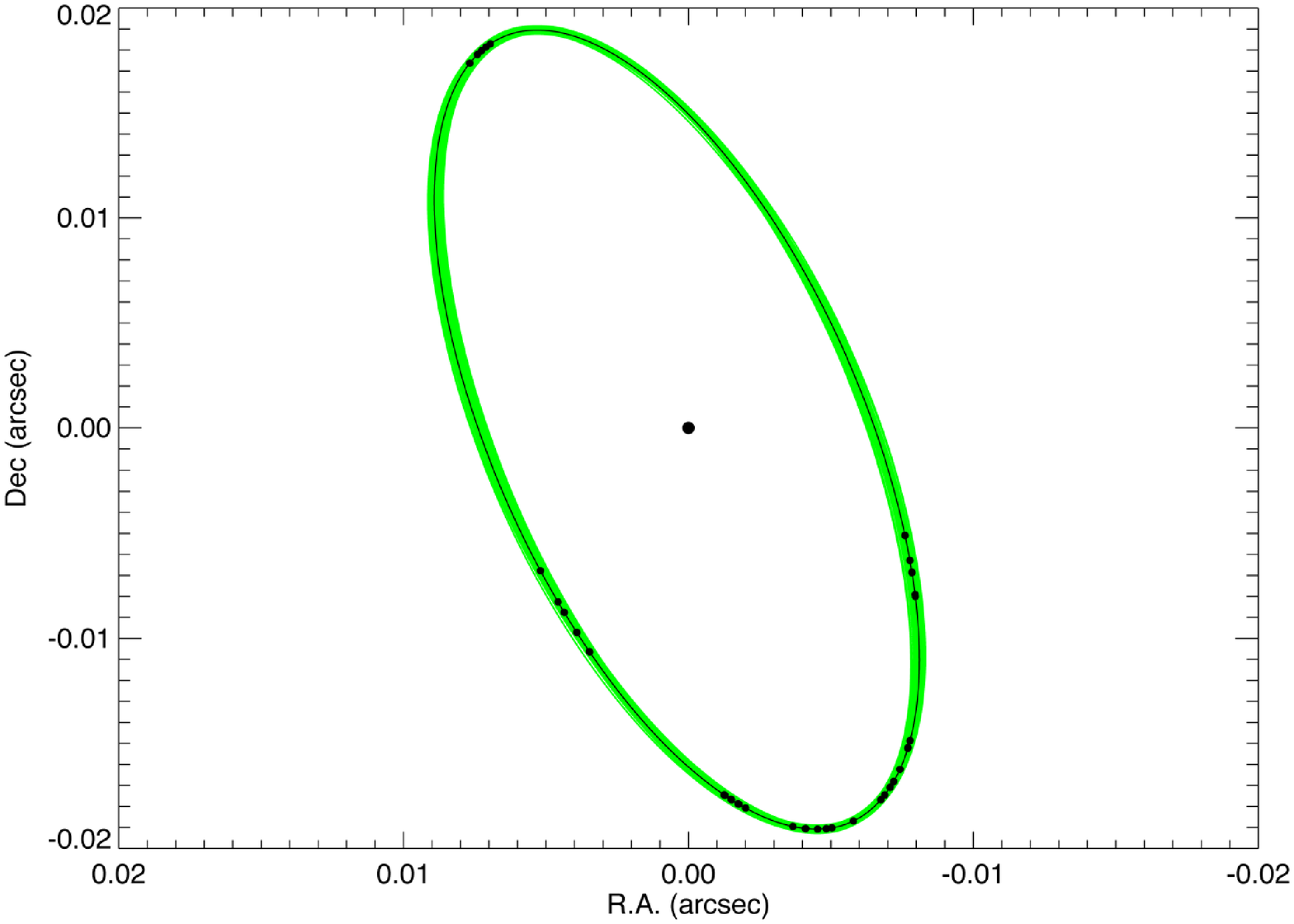}}
\rotate
\hphantom{.....}
\caption{
The best-fit visual orbit of \lamvir. The primary
is shown by the central dot. The solid line indicates the best-fit orbit,
and the overplotted filled dots show the epochs of interferometric observations.
The shaded area around the orbit indicates the 1-$\sigma$ uncertainties of the orbit.
\label{orbit}}
\end{center}
\end{figure}

The preliminary orbital parameters are shown in the third column of
Table~\ref{bestfit}.
As can
be seen in the table, the $\chi^2_{\nu}$ of the squared-visibilities and closure phases
are still larger than unity. In fact, these large $\chi^2_{\nu}$s are due to
the systematic bias in the closure phases caused by the bandwidth smearing
effect mentioned previously and also in the Appendix, which tends to change the flux ratio and
cannot be eliminated by the new estimated errors. In order to reduce this bias
 and other uncertainties in the closure phases, we
conservatively give small weight to the closure phases in the fit
such that the orbital parameters
primarily come from the squared-visibilities and the radial velocities.
The weight for the closure phases is determined iteratively in the fit until the deweighting of closure 
phases does not change the flux ratio any more.
Fig.~\ref{orbit} depicts the best-fit visual orbit of \lamvir, and
the final best-fit parameters are listed in the fourth column of
Table~\ref{bestfit}. The de-weighting of the closure phases also improved the
$\chi^2_{\nu}$ of the visibilities, as well as the overall fit. The value of the
flux ratio increased a significant amount due to the elimination of the
closure phase bias.
For reference, we also list the parameters from \citet{Stickland1975} in the table.
Due to the measurement
uncertainties of \cite{Stickland1975} and the near-equal masses of the two
components, the primary and secondary
components were reversed, resulting in a 180$\arcdeg$
difference in $\omega$ compared to our value.
We have corrected this in Table~\ref{bestfit}.

\section{Physical parameters}
\label{sec:physics}

The combination of the astrometric and spectroscopic information
provided by our orbital solution yields precise estimates of the
absolute masses of the components of \lamvir, with relative errors of
only 0.7\% for the primary and 1.5\% for the secondary. These are
listed in Table~\ref{tab:physics} along with other physical parameters
described below. We use these in the next section to compare against
recent stellar evolution models and assess the evolutionary state of
the system.

\begin{figure}[thb]
\begin{center}
{
\includegraphics[angle=0,width=6in]{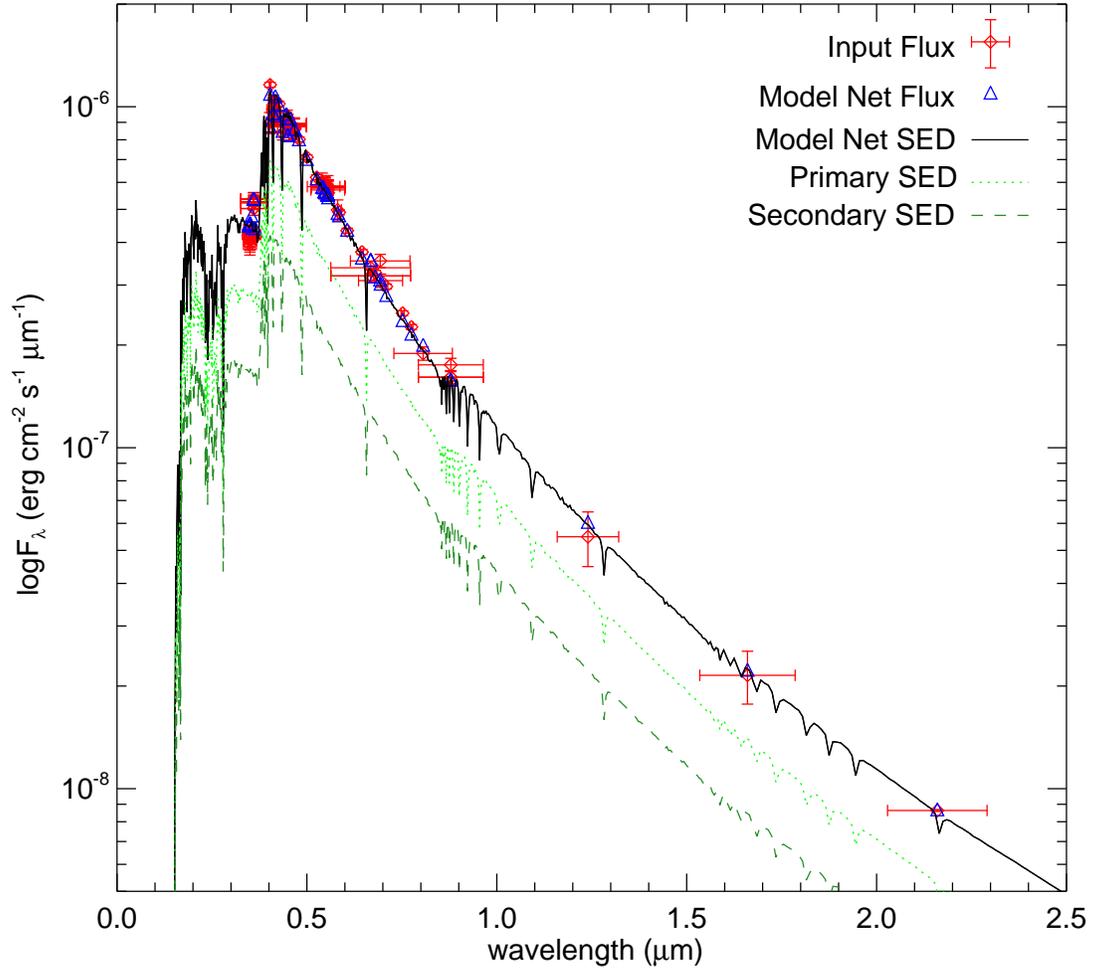}}
\rotate
\hphantom{.....}
\caption{
SED models of \lamvir. The net SED model is shown by the solid line, overplotted
with input flux and the bandpass integrated model flux. The bandpass of input fluxes
are shown by the horizontal error-bars. The SED for
the primary is shown by the dotted line and the
secondary by the dashed line. The models correspond to two A1V stars.
\label{sed}}
\end{center}
\end{figure}

The system bolometric flux and luminosities are determined through spectral energy distribution
(SED) modelling. We constructed two-component SED models using both Kurucz and
Pickles model templates \citep{Kurucz1974, Pickles1998} and applied them to a substantial
amount of archive photometric measurements in the Johnson, Str\"omgren, Geneva
 and 2MASS systems, as well as spectrophotometric
measurements from \citet{Breger1976}, \citet{Burnashev1985},
and \citet{Glushneva1998}. However, the \cite{Burnashev1985} and \cite{Glushneva1998} spectrophotometry
 are not consistent with photometry at wavelengths longer than 420nm, and the \cite{Burnashev1985} 
data also have bad normalizations that do not agree with other data. Therefore, we only take the spectrophotometry of \citet{Breger1976} into account in our fitting. The component
light ratios determined from spectroscopy and interferometry in
\S\ref{obs} are also used to constrain the models. After extensive
tests of model templates, we found that the Pickles templates are
not appropriate for \lamvir\ because of its metallic-lined
nature. The Kurucz model with [m/H] = +0.5 best fits the data.
Fig.~\ref{sed} depicts the resulting best-fit Kurucz model, as well as the
corresponding SEDs for both the primary and the secondary, overplotted
with the input flux measurements and the model net flux for corresponding
bandpasses. The best-fit model calls for two A1V stars 
with no need of extinction correction. The resulting system bolometric flux is
$3.794\times10^{-7} \pm 0.014\times10^{-7}$ erg cm$^{-2}$
s$^{-1}$. With the distance determined below, the luminosities for
the primary and the secondary are $20.84 \pm 0.25$ \lsun~ and
$12.58 \pm 0.16$ \lsun~respectively.

The effective temperature estimates in \S\ref{sec:spectroscopy}
are strongly correlated with the metallicity adopted for \lamvir,
in the sense that higher metallicities lead to higher
temperatures. Consequently, because the composition in the surface
layers of \lamvir\ is enhanced compared to normal A stars, our
temperatures are likely to be overestimated. We therefore made use
of extensive photometric measurements available for the object in
the Johnson, Str\"omgren, and Geneva systems \citep{Mermilliod:97}
as well as 2MASS, to derive the mean effective temperature based
on a large number of color/temperature calibrations
\citep{Popper:80, Moon:85, Blackwell:90, Gray:92, Napiwotzki:93,
Balona:94, Smalley:95, Kunzli:97, Cox:00}. In addition we made an
estimate by the infrared flux method \citep{Blackwell:90} based on
the bolometric flux determined from the SED, the flux from the
2MASS $K_s$ band, and the corresponding integrated flux from the
Kurucz model. The various estimates are in good agreement, showing
a scatter of about 120~K and yielding an average of $T_{\rm eff} =
8280 \pm 200$~K, where the uncertainty is a conservative estimate
to account also for the possibility of systematics errors in the
calibrations. To the extent that the abundance enhancement of the
two stars is similar \citep[which appears to be the case, as
reported by][]{Stickland1975}, our spectroscopic analysis in
\S\ref{sec:spectroscopy} indicates no significant difference in
temperature between the stars.  Reddening estimates based on
Geneva and Str\"omgren photometry give negligible values using
calibrations by \cite{Crawford:79} and \cite{Kunzli:97},
consistent with the relatively close distance to the object.

The orbital parallax of the system is $\pi_{\rm orb} = 18.81 \pm
0.10$ mas, corresponding to a distance of $53.16 \pm 0.29$ pc. The
trigonometric parallax listed in the {\it Hipparcos\/} catalog is
$\pi_{\rm HIP} = 17.47 \pm 0.94$ mas, which is slightly lower than
ours (a 1.4$\sigma$ or 7\% effect) most likely because it does not
account for the perturbation from the orbital motion. 
The original {\it Hipparcos\/} observations are available in the form
of `abscissa residuals', which are the one-dimensional residuals
(along the scan direction of the satellite) from the usual 5-parameter
solutions yielding the position, proper motion, and parallax as
reported in the Catalogue (ESA 1997). We have re-reduced these
measurements by expanding the model to account for the orbital motion
constrained using our own solution, and we have solved for the
semimajor axis of the photocenter ($a_{\rm phot}$) as well as
corrections to the position and proper motion of the barycenter and a
correction to the parallax. The formalism for this solution follows
closely that described by \cite{vanLeeuwen:98} and \cite{Pourbaix:00},
and a recent example of a similar application is given by
\cite{Torres:06}. The revised {\it Hipparcos\/} parallax we obtain is
$\pi_{\rm HIP}^{\prime} = 18.55 \pm 0.84$ mas, which is now in much
better agreement with $\pi_{\rm orb}$ (within 0.3 $\sigma$). The
motion of the center of light of the binary is clearly detected by
{\it Hipparcos}, albeit with much lower precision than the relative
semimajor axis, and amounts to $a_{\rm phot} = 3.84 \pm 0.63$
mas\footnote{For completeness we list here the remaining parameters
adjusted in this fit: $\Delta\alpha\cos\delta = -0.19 \pm 0.77$ mas,
$\Delta\delta = +0.03 \pm 0.47$ mas, $\Delta\mu_{\alpha}\cos\delta =
+1.48 \pm 0.81$ mas~yr$^{-1}$, and $\Delta\mu_{\delta} = +0.47 \pm
0.58$ mas~yr$^{-1}$. These should be added with their sign to the
catalog values of the position and proper motion to yield the revised
values.}.

This value along with the relative semimajor axis and the mass
ratio allow us to obtain an independent estimate of the light
ratio in the {\it Hipparcos\/} passband ($H_p$), which is
$\ell_2/\ell_1 = 0.39 \pm 0.06$. This is significantly lower than
the spectroscopic and interferometric value in \S\ref{obs}. There
is no evidence from either the spectroscopy or the interferometry
of any photometric variability in \lamvir\ that might explain the
difference, in agreement with the small scatter observed in the
brightness measurements from {\it Hipparcos\/}
\citep[$\sigma_{H_p} = 0.006$ mag;][]{ESA:97}. The small amplitude
of the photocentric motion compared to the median error of an
individual abscissa residual (2.3 mas) may be cause for some
concern about possible systematics in the {\it Hipparcos\/} light
ratio, although we have no independent evidence for such an
effect. On the other hand, in view of the metallic-lined nature of
the stars we cannot entirely rule out the possibility of a bias in
the spectroscopic value of $\ell_2/\ell_1$ of a similar nature as
the effect in the temperatures mentioned above. However, the
brightness ratio is a differential measurement and therefore we
would not expect the effect to be large.  Since both light ratio
estimates are close to the visual band and the stars are of
similar temperature, for the purpose of the modelling in the next
section we have chosen as a compromise to adopt the weighted
average of the spectroscopic and {\it Hipparcos\/} values.  That
average is $\ell_2/\ell_1 = 0.56 \pm 0.10$. The larger uncertainty
accounts for the individual weights as well as the difference in
the values themselves.

The absolute visual magnitudes of the components follow from this
value along with the orbital parallax and the apparent system
magnitude of $V = 4.52 \pm 0.01$ \citep{Mermilliod:97}, and are
included in Table~\ref{tab:physics}. Although we have derived a very
precise flux ratio in the $H$ band from the interferometric
observations, a total $H$-band magnitude for the system is unavailable
(the star is bright enough that it saturated the 2MASS detector), and
so the individual magnitudes cannot be computed directly.

\section{Comparison with stellar evolution models}
\label{sec:models}

The accurately measured masses, absolute visual magnitudes, and
effective temperatures of the components of \lamvir, as well as the
flux ratio in the $H$ band, allow a comparison with current models of
stellar evolution.  For their ease of use we have chosen here the
Yonsei-Yale series of calculations by \cite{Yi:01} and
\cite{Demarque:04}. The color/temperature transformations and
bolometric corrections adopted are those of \cite{Lejeune:98}, and the
passband of the $H$ filter in those calculations is sufficiently close
to that used at IOTA for our purposes. Unfortunately the actual bulk
composition of \lamvir\ is difficult or impossible to determine
observationally because of the metallic-lined nature of both
stars. Therefore, we have explored a range of interior metallicities
in the models to identify the values that are consistent with the
observations.

\begin{figure}
\vskip -1.5in
\epsscale{1.0}
\plotone{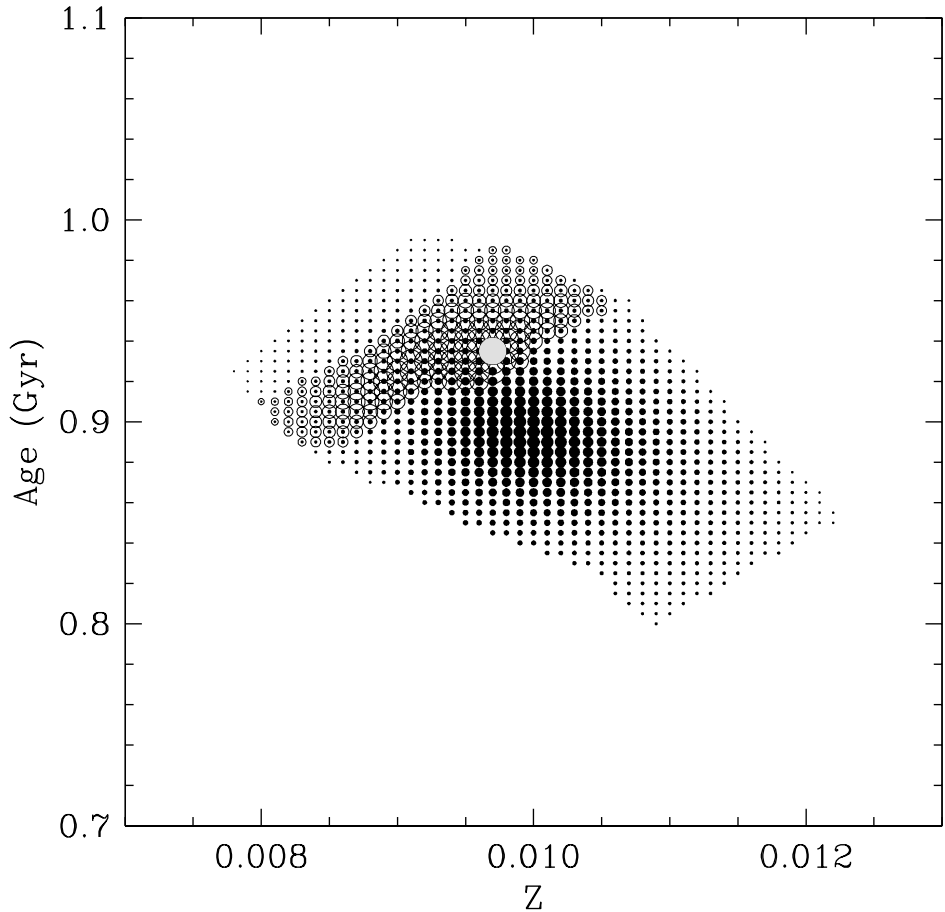}
\vskip -1.5in

 \figcaption[]{Determination of the age and metallicity of \lamvir\ by
 comparison with stellar evolution models by \cite{Yi:01} and
 \cite{Demarque:04}. Filled circles show all age/metallicity
 combinations that yield an isochrone matching the measured values of
 the mass, effective temperature, and absolute visual magnitude of
 both stars (assumed to be coeval) within the observational errors
 (Table~\ref{tab:physics}). Larger filled circles indicate a better
 match. Open circles indicate age/metallicity combinations that in
 addition satisfy the measured flux ratio in the $H$ band, within its
 uncertainty. The size of the open circles is again proportional to
 the goodness of fit. The best overall match to the observations is
 indicated with the large grey circle, and corresponds to $Z = 0.0097$
 (or [Fe/H] = $-0.29$) and an age of 935~Myr. \label{fig:agemet}}

 \end{figure}

Initially we considered only the masses, absolute magnitudes, and
effective temperatures of the two stars as constraints. By
interpolation we computed a fine grid of isochrones for a large number
of age and metallicity combinations, and compared each model with the
6 measurements under the assumption that the stars are coeval and have
the same interior composition. The result is shown in
Fig.~\ref{fig:agemet}, where each filled circle represents an
age/metallicity combination that agrees with the observations for both
stars within the errors. The best agreement occurs near the center of
the region (larger filled circles), at a metallicity near $Z = 0.01$
and an age of about 900~Myr. Next we added the constraint on the flux
ratio in $H$, requiring that in addition to matching the masses,
magnitudes, and temperatures, the models reproduce the observed
magnitude difference in $H$. The combinations that also satisfy this
last constraint cover a smaller area of the diagram, and are indicated
with open circles in Fig.~\ref{fig:agemet}. The best overall match
is achieved for a metallicity of $Z = 0.0097$ (corresponding to [Fe/H]
= $-0.29$, assuming no enhancement of the alpha elements) and an age
$t$ of 935~Myr, at which the models agree with all observables well
within the errors (typically to better than 0.4$\sigma$). 

\begin{figure}
\vskip -0.5in
\epsscale{0.90}
\plotone{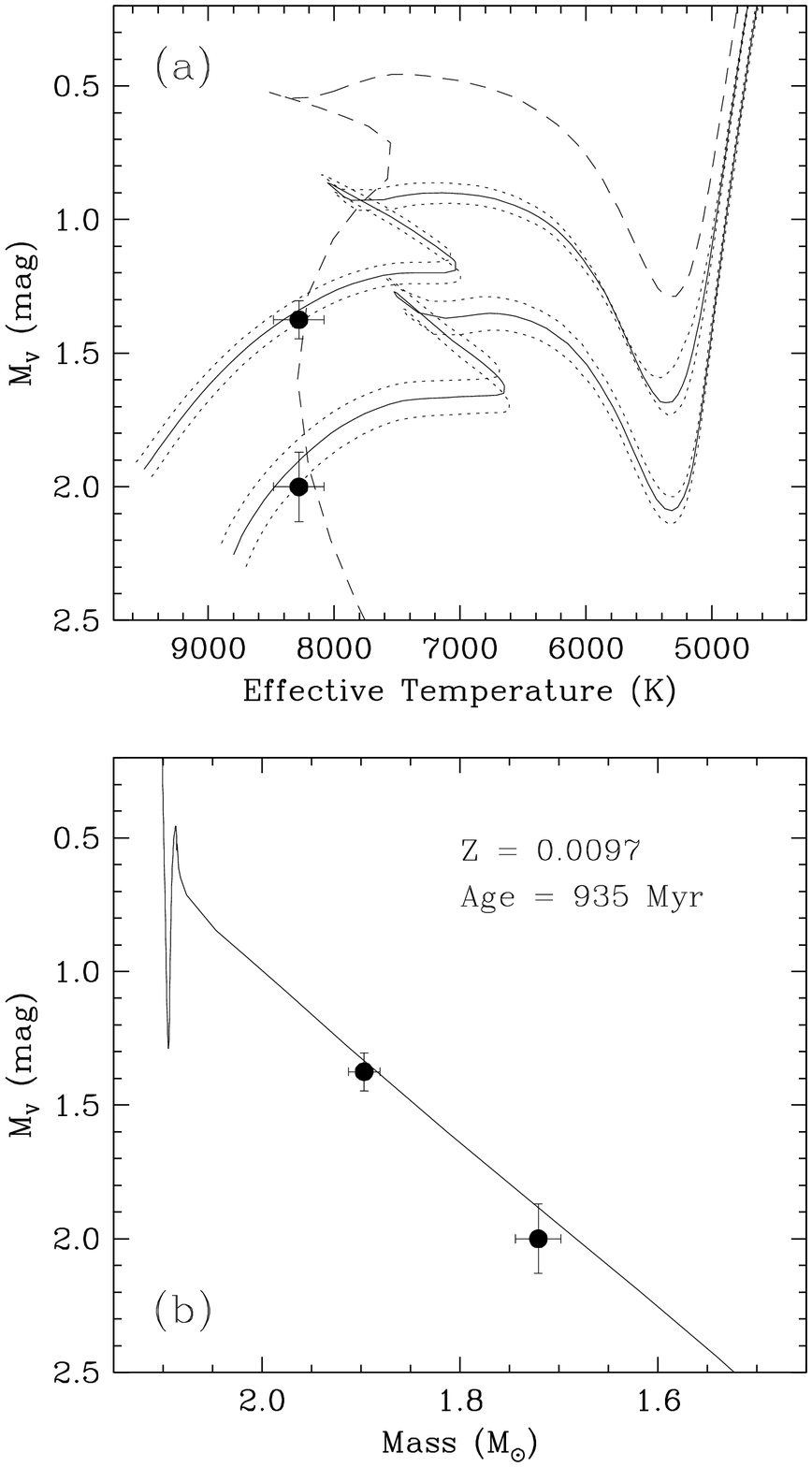}
\vskip -0.4in

 \figcaption[]{Comparison between the measurements for \lamvir\ and
 the best matching stellar evolution models by \cite{Yi:01} and
 \cite{Demarque:04}, for a metallicity of $Z = 0.0097$ (or [Fe/H] =
 $-0.29$) and an age of 935~Myr. (a) Evolutionary tracks in the
 absolute magnitude/effective temperature diagram for the exact masses
 measured for each star (solid lines). The uncertainty in the location
 of the tracks stemming from the mass errors ($\pm$1$\sigma$) is
 represented with the dotted lines. The 935-Myr isochrone is shown by
 the dashed line. (b) Best-fitting isochrone in the mass-luminosity
 diagram. \label{fig:tracks}}

 \end{figure}

\begin{figure}
\vskip -1.5in
\epsscale{1.0}
\plotone{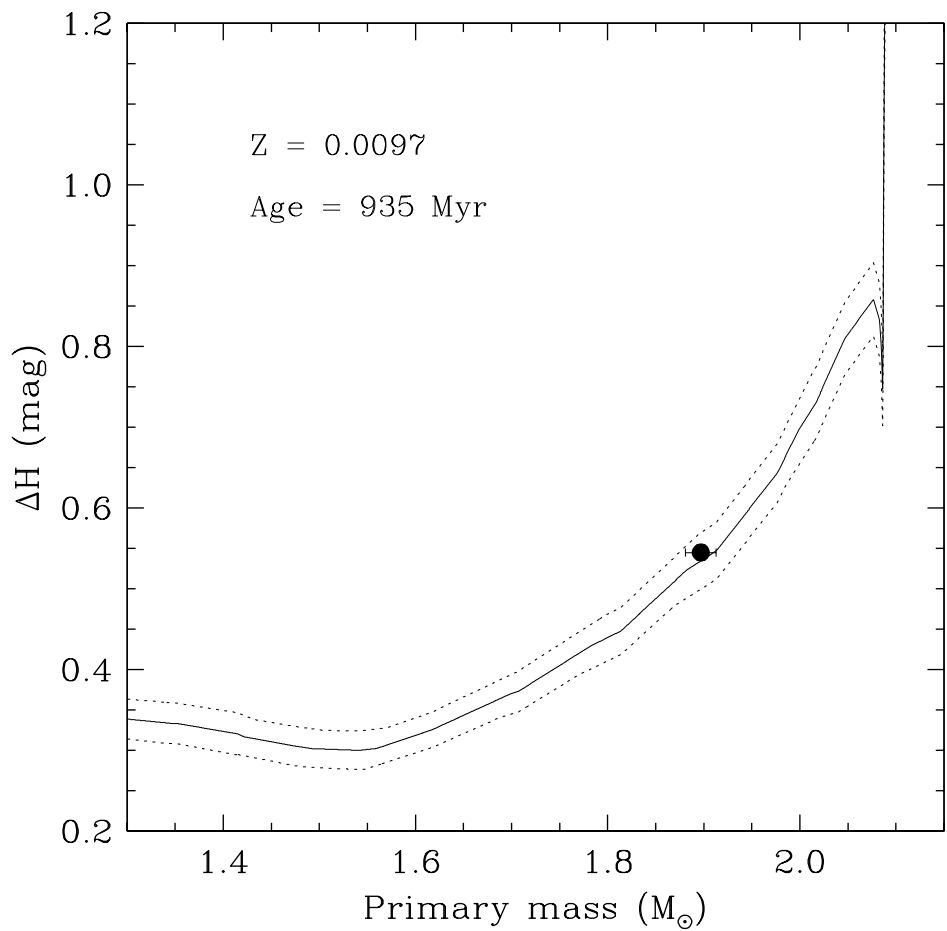}
\vskip -2.0in

 \figcaption[]{Predicted magnitude difference in the $H$ band from the
 best-fitting model isochrone for \lamvir, compared with our accurate
 measurement from IOTA (vertical error bar smaller than the size of
 the point). The solid line is the prediction for the exact mass ratio
 $q$ we measure. At each point along this line the secondary mass is
 computed from the primary mass and $q$, and the magnitude difference
 read off from the isochrone. The dotted lines represent the
 uncertainty in the prediction resulting from the error in
 $q$. \label{fig:fluxratio}}

 \end{figure}

The comparison of the masses, absolute magnitudes, and temperatures
with the models is shown graphically in Fig.~\ref{fig:tracks}. The
solid lines in the top panel represent evolutionary tracks computed
for the exact masses we measured for each star, and the dotted lines
indicate the uncertainty associated with the mass errors
($\pm$1$\sigma$). The 935-Myr isochrone is shown as a dashed line, and
indicates that the components of \lamvir\ are indeed consistent with
having the same age, as expected. Fig.~\ref{fig:tracks}b shows the
best-fit model isochrone and the observations in the mass-luminosity
diagram. The constraint on the flux ratio is illustrated in
Fig.~\ref{fig:fluxratio}, where we have chosen to represent the
predicted magnitude difference from the model (solid line) as a
function of the primary mass, with the secondary mass being determined
at each point along the curve from the measured mass ratio ($q \equiv
M_2/M_1$). The dotted lines represent the uncertainty in the location
of this curve ($\pm$1$\sigma$) resulting from the error in $q$. The
measurement is in good agreement with the predictions.

The estimated radii of the stars from the best fitting model are $R_1
= 2.35$~R$_{\sun}$ and $R_2 = 1.84$~R$_{\sun}$ for the primary and
secondary, respectively, and the corresponding angular diameters at
the distance of \lamvir\ are $\phi_1 = 0.41$~mas and $\phi_2 =
0.32$~mas. These are not far from the values adopted for the orbital
solution described in \S\ref{finalorb}. The surface gravities are
$\log g_1 = 3.97$ and $\log g_2 = 4.14$, which are close to the value
of $\log g = 4.0$ adopted for both components in
\S\ref{sec:spectroscopy}.

\section{Comparison with tidal theory}
\label{sec:tidal}

The measures of the absolute dimensions as well as the projected
rotational velocities $v \sin i$ of the components of \lamvir\ allow
us to test various aspects of tidal evolution theory. Tidal forces in
binaries tend to synchronize the rotation of each star to the mean
orbital motion, to align the spin axes of the stars with the axis of
the orbit, and also to circularize the orbit.

In general the timescales for these processes are very different
\citep[see, e.g.,][]{Hut:81}. 
Alignment and synchronization typically proceed much more quickly
than circularization, often by an order of magnitude or more when the
angular momentum of the orbit is larger than the rotational angular
momentum. Tidal forces are highly
sensitive to the dimensions and structure of the stars.  Both
components of \lamvir\ started their main-sequence lives with
convective cores and radiative envelopes, but in later evolutionary
stages their envelopes will become convective.  Therefore, it is
necessary to consider two different mechanisms of tidal braking
appropriate for each stage, which are referred to as radiative damping
and turbulent dissipation, respectively \citep{Zahn:77, Zahn:89}. The
timescales for synchronization and circularization for the case of
stars with convective envelopes are given by
 $${\tau_{\rm sync}}= 3.95\times10^2 \beta^2
{M^{7/3}}{{(1+q)^2}\over{q^2}}
{L^{-1/3}}{{\lambda_{2}}^{-1}}{P^{4}\over{R^{16/3}}}$$
 $${\tau_{\rm circ}}= 1.99\times10^3
{M^{3}}{{(1+q)^{5/3}}\over{q}}{L^{-1/3}}
{{\lambda_2}^{-1}}{P^{16/3}\over{R^{22/3}}}~,$$
while for stars with convective cores and radiative envelopes the
timescales are
 $${\tau_{\rm sync}}= 2.03 \beta^2 {M^{7/3}}{{(1+q)^2}\over{q^2}}{{E_2}^
{-1}}{P^{17/3}\over{R^{7}}}$$
 $${\tau_{\rm circ}}= 1.71\times10^1 {M^3}{{(1+q)^{5/3}}\over{q}}
{{E_2}^{-1}}{P^{7}\over{R^9}}~.$$
In the above expressions the timescales are given in years, $q$
represents the mass ratio, and $M$, $R$ and $L$ are the mass, radius,
and luminosity in solar units.  The period $P$ is given in days. The
symbol $\lambda_2$ represents the tidal coefficient
\citep[see][]{Zahn:89}, $\beta$ is the fractional radius of gyration,
and the coefficient $E_2$ is related to the dynamical tidal
contribution to the total perturbed potential \citep[see][Eq.(6) and
following]{Claret:97}.  In order to consider the contribution of both
components to the circularization we use the equivalent time scale
 $$-\frac{1}{e}\frac{de}{dt}= \frac{1}{\tau_{\rm circ,1}} +
\frac{1}{\tau_{\rm circ,2}}~,$$
where subscripts 1 and 2 refer to the primary and the secondary
components. The differential equations that govern the evolution of
the eccentricity and axial rotation were integrated along evolutionary
tracks for each star until the relative variations reached 0.05\% of
their initial values. The integrations were carried out using the
fourth-order Runge-Kutta method.

\begin{figure}
\vskip -1.0in
\epsscale{1.0}
\plotone{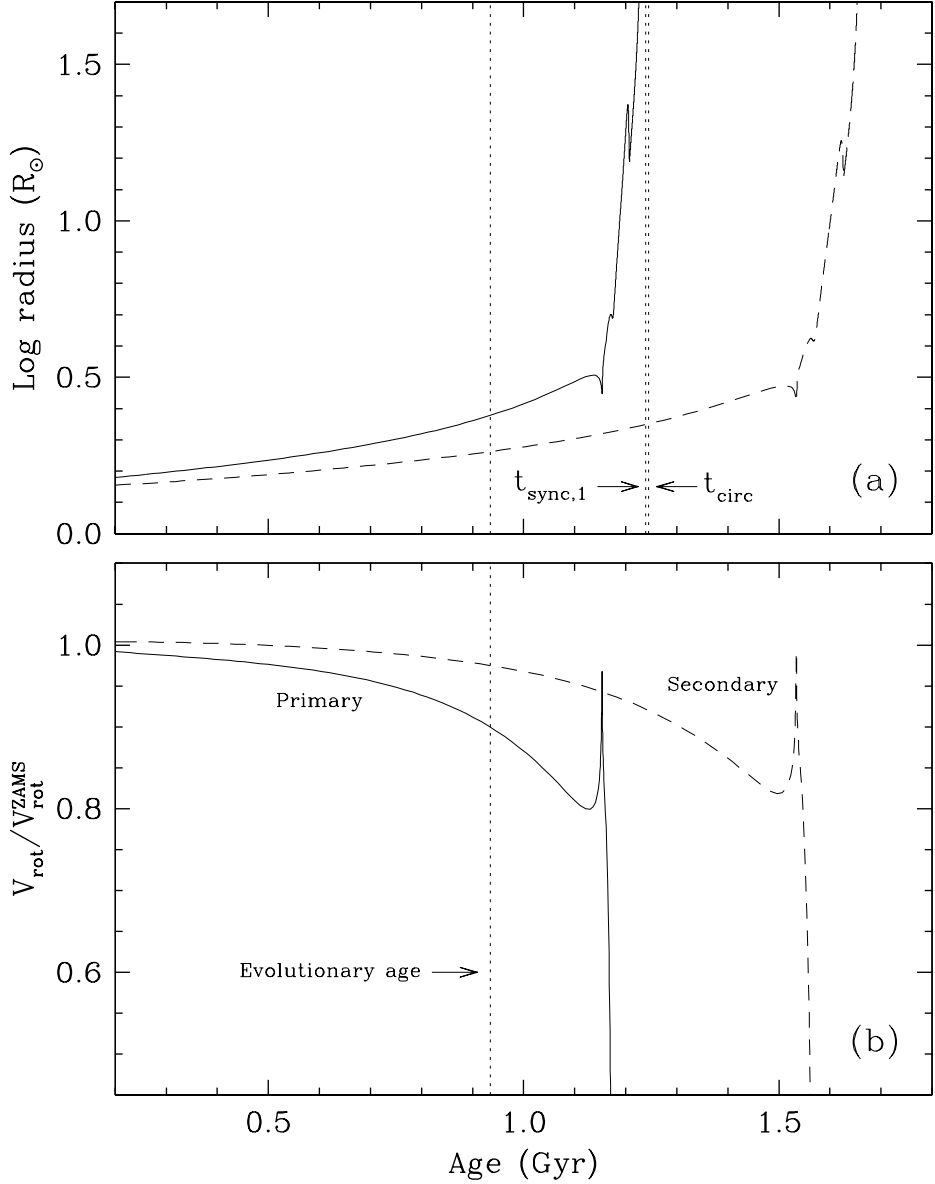}
\vskip -1.2in

 \figcaption[]{Tidal evolution of \lamvir.
(a) Expected change in radius as a
 function of time based on the models by \cite{Claret:04}. The present
 evolutionary age of the binary is indicated, along with the predicted
 times of synchronization of the primary and of circularization of the
 orbit due to tidal forces.  (b) Evolution of the
 rotational velocity of each component relative to the initial
 rotation upon arrival on the ZAMS, due solely to the change in the
 moment of inertia with time. \label{fig:tidal}}
 \end{figure}

We find that the time of circularization of the orbit is predicted to
be $t_{\rm circ} = 1.245$~Gyr, which is larger than the present age of
the system. Thus the non-zero eccentricity we measure for the orbit of
\lamvir\ ($e = 0.0610 \pm 0.0036$) is consistent with theory. The
primary star is expected to become synchronized with the mean orbital
motion at a slightly earlier time $t_{\rm sync,1} = 1.239$~Gyr.  Once
again this agrees with theoretical expectations, since the synchronous
velocity of the star at the present time would be well under 1~\kms,
whereas we measure $v_1 \sin i = 36$~\kms. The times $t_{\rm circ}$
and $t_{\rm sync,1}$ are indicated with vertical dotted lines in
Figure~\ref{fig:tidal}a, which depicts the evolution of the radius of
the two stars as a function of age, for reference. The evolutionary
age of the system (0.935~Gyr; \S\ref{sec:models}) is also
indicated. It is seen that synchronization of the primary and
circularization of the orbit are triggered by the relatively sudden
increase in size suffered by the primary as it moves up the giant
branch. For $t_{\rm sync,2}$ we can only place a lower limit of
1.245~Gyr (the same as $t_{\rm circ}$) because the nuclear timescale
of the secondary is 35\% longer than the primary, and the evolutionary
tracks do not reach sufficiently advanced stages to allow the
integrations. This is again consistent with the fact that the measured
$v \sin i$ of the secondary (10~\kms) is much larger than the
synchronous value (which is similar to the primary).
                                                                                
One of the characteristics of the \lamvir\ system that has drawn
attention in the past, particularly in connection with the Am nature
of the binary, is the \emph{difference} in the projected rotational
velocities of the components \citep[see, e.g.,][]{Stickland1975}. From
our measurements in \S\ref{sec:spectroscopy} the primary is rotating
approximately 3.5 times more rapidly than the secondary. Both values
of $v \sin i$ are low compared to the average for A-type stars in the
field, which has typically been found to be the case for all Am stars.
Since these objects are overwhelmingly found to be members of binary
systems \citep{Abt1961}, the connection between the slow rotation, the
chemical peculiarities, and binarity has been much discussed
\citep[see, e.g.,][]{Abt1995, Budaj:96, Budaj:97, Abt2000,
Bohm-Vitense:06}.  In the following we examine extent to which
differences in the evolution of the spin rates and/or differences in
the evolution of the orientation of the rotation axes since formation
might have contributed to the difference in the $v \sin i$ values
presently observed.
                                                                                
The evolution of the spin of each star is a function of the changes in
the moment of inertia due to evolution, and also depends on the
effects of tidal forces from the companion. As described above, the
latter are expected to be relatively weak in the present evolutionary
state of the binary, since synchronization is not expected to happen
for another 300 Myr ($\sim$1/3 of the present age of \lamvir).  To
model the changes in the moment of inertia we have made use of the
Granada series of stellar evolution calculations \citep{Claret:04}
that are well-suited for binary studies. The physics in these models
is similar to that in the Yonsei-Yale models, though some of the
details are somewhat different. We adopt a composition similar to that
found earlier ($Z = 0.01$), and we computed mass tracks for the exact
masses we measure for the stars.  In Figure~\ref{fig:tidal}b we show
the change in the rotational velocity of each component with time
relative to its initial value on arrival at the ZAMS. At the present
evolutionary age of the system the rotational velocity of the primary
is predicted to have slowed by about 10\% from its initial value,
whereas the decrease for the secondary is only 2--3\%.  Given that the
$v \sin i$ of the primary is currently the higher of the two, an
obvious possibility is that the primary was initially rotating more
rapidly than the secondary. Another is that the spin axes of the stars
have different inclinations relative to the line of sight.
                                                                                
As indicated earlier the timescale for alignment of the rotation axes
of the binary components with the axis of the orbit due to tidal
forces is typically much shorter than the timescale for
circularization, and as a result alignment is virtually always
assumed. To examine whether this is actually true for \lamvir\ we
consider, in addition to the differential equations of tidal evolution
used above, one that describes the evolution of the orientation of the
spin axis of each star, characterized by an inclination angle $i$.  We
follow closely the formalism by \cite{Hut:81}, with a timescale for
the alignment of the spin axis with the axis of the orbit given in
terms of the circularization timescale by
 $${\tau_{i}}= \frac{7} {(\alpha + 1)} {\tau_{\rm circ}}~.$$
In this expression $\alpha$ is the ratio between the orbital and
rotational angular momenta, which is given by
 $$\alpha = \frac{q}{1+q} \left(a\over{\beta R}\right)^2~,$$
where $a$ is the semimajor axis of the orbit. The integrations were
carried out as described above. The expected time for the alignment of
the spin axis of the primary is 1.239~Gyr, not surprisingly the same
as the time of synchronization. For the secondary once again we can
only place a lower limit of 1.245~Gyr. It follows that neither of the
rotation axes is expected to be aligned with the orbit, unless they
were perfectly aligned to begin with. Therefore, the projection factor
for the equatorial rotational velocities remains unknown (i.e., it is
not necessarily the same as the inclination of the orbit, which we
have measured precisely) and may be different for each star.
                                                                                
In conclusion, the measured difference in the projected rotational
velocities of \lamvir\ may result from the very different initial
rotation rates of the two components, the different projections of
their equatorial velocities due to misalignment of the spin axes, or
perhaps to the combination of the two effects.

\section{Summary and conclusion}
\label{sec:summary}

By combining the interferometric and radial velocity data, we have
determined the 3-dimensional orbital of \lamvir\ to high precision.
We studied the effects of bandwidth smearing on squared-visibilities
and closure phases.
The calibration problems in the closure phases due to these effects are
larger than expected, suggesting the necessity of using narrow bandwidth
for precision work.

Our precise determination of the \lamvir\ orbit allows us to study
its physical properties accurately. We determined the masses of
the two components with accuracies of $0.7\%$ and $1.5\%$ for the
primary and the secondary respectively. We studied the SED of
\lamvir\ with archival photometric and spectrophotometric data. The
Kurucz model templates with [m/H] = +0.5 fit the data best, yielding
a solution with two A1V type stars. The temperatures of the
system are derived from various methods, leading to a value of
8280$\pm$200 K for both stars. Other properties of \lamvir, such
as distance, bolometric flux, luminosity, radii, motion of the
photocenter, are also determined.

The accurately determined properties allow a comparison with
current stellar evolution models. The model that matches best
yields a subsolar metallicity of $Z=0.0097$ and an age of 935 Myr,
indicating the evolution of \lamvir\ is similar to normal A stars
despite their surface abundance anomalies. A study of tidal
evolution in \lamvir\ indicates that its orbital circularization
time is $t_{circ}=1.245$ Gyr, larger than the present age of the
system, and therefore theory agrees with the observation that the orbit is not
currently circular. The predicted orbital synchronization time also
implies that neither of the two stars has synchronized rotation.
Furthermore, neither of the rotational axes is expected to be
aligned with the orbit, implying that in addition to the
possibility that the two stars have very different initial
rotations, the measured differing rotational velocities may also
stem from the projection of the equatorial rotational velocities.

The origin of the abundance anomalies of Am stars has been a
puzzle for a quite some time. It is widely believed that when stars are
slow rotators, atomic diffusion will play an important role in the
outer convection zones, causing abnormal abundances and therefore
the Am phenomenon. However, different views have also been presented \citep{Bohm-Vitense:06}.
Although Am stars have been studied
intensively, only a few of them have well determined properties.
With its accurately determined physical properties and well known
evolution status, as well as its possibly differing rotation rates
(which may lead to different diffusion efficiencies), 
\lamvir\ is an ideal candidate for follow up studies such as
detailed abundance analyses and atomic diffusion modelling that
can shed light on our understanding of the causes of the Am
phenomenon.

\acknowledgments
We thank C.\ R.\ Cowley for helpful discussion and advice, and J.\
R.\ Caruso, R.\ J.\ Davis, D.\ W.\ Latham, T.\ Mazeh, A.\ A.\ E.\
Milone, R.\ P.\ Stefanik, J.\ M.\ Zajac for obtaining most of the
spectroscopic observations used in this work. 
We also thank the referee for a number of helpful comments.
We gratefully
acknowledge support for IOTA from the Smithsonian Astrophysical
Observatory, the National Aeronautics and Space Administration (for
third telescope development, NNG05G1180G; for data analysis support,
NNG04GI33G), and the National Science Foundation (AST-0138303,
AST-0352723). GT wishes to acknowledge partial support for this work
from NSF grant AST-0406183 and NASA's MASSIF SIM Key Project
(BLF57-04). AB also gratefully acknowledges the support of NASA. The IONIC3
instrument has been
developed by LAOG and LETI in the context of the IONIC collaboration (LAOG, IMEP,
LETI).
The IONIC project is funded by the CNRS and CNES (France).
 This research has made use of the SIMBAD database, operated
at CDS, Strasbourg, France, and NASA's Astrophysics Data System Abstract
Service. This publication makes use of data products from the Two Micron
All Sky Survey, which is a joint project of the University of
Massachusetts and the Infrared Processing and Analysis Center/California
Institute of Technology, funded by NASA and NSF. This work also makes use
of services produced at the Michelson Science Center at the California
Institute of Technology.

\bibliographystyle{apj}  
\bibliography{apj-jour,lamvir}

\clearpage


\begin{deluxetable}{lcll}
\footnotesize
\tabletypesize{\scriptsize}
\tablecaption{Radial Velocity of \lam Vir
\label{tab:rvs}}

\tablehead{
  \colhead{HJD} & \colhead{Orbital Phase} & \colhead{$RV_1$ (\kms)} & \colhead{$RV_2$ (\kms)}}
\startdata
45156.5547  &  0.72  &  -31.80  & 18.58  \\
46576.6328  &  0.59  &  -19.43  &  5.80  \\
46576.6758  &  0.59  &  -20.03  &  5.50  \\
46576.6797  &  0.59  &  -19.24  &  5.70  \\
46597.6406  &  0.69  &  -30.60  & 16.41  \\
46597.6562  &  0.69  &  -30.60  & 16.34  \\
46597.6641  &  0.69  &  -31.93  & 17.30  \\
46611.5703  &  0.76  &  -33.75  & 19.10  \\
46613.5664  &  0.77  &  -32.03  & 19.33  \\
46633.6016  &  0.86  &  -26.78  & 13.07  \\
46635.5430  &  0.87  &  -27.06  & 11.62  \\
46636.5430  &  0.88  &  -28.03  & 10.66  \\
46640.5586  &  0.90  &  -23.66  &  9.12  \\
46809.9688  &  0.72  &  -32.51  & 18.19  \\
46819.9570  &  0.77  &  -34.08  & 19.31  \\
46896.7656  &  0.14  &   14.56  &-30.78  \\
46918.7227  &  0.24  &   17.31  &-34.79  \\
46924.6719  &  0.27  &   16.75  &-33.91  \\
46938.6758  &  0.34  &   12.25  &-28.73  \\
46953.6094  &  0.41  &    1.66  &-20.67  \\
47197.9453  &  0.59  &  -19.45  &  6.86  \\
47206.0508  &  0.63  &  -25.53  & 11.80  \\
47218.9141  &  0.70  &  -30.68  & 18.07  \\
47222.8750  &  0.71  &  -32.75  & 16.99  \\
47226.8438  &  0.73  &  -30.44  & 20.05  \\
47320.7148  &  0.19  &   17.45  &-34.27  \\
47568.8828  &  0.39  &    5.82  &-23.65  \\
47569.9688  &  0.39  &    5.08  &-22.15  \\
47570.9414  &  0.40  &    2.87  &-22.46  \\
47574.9297  &  0.42  &    0.86  &-20.01  \\
47575.9141  &  0.42  &   -0.60  &-19.07  \\
47583.8047  &  0.46  &   -6.53  &-13.33  \\
47585.9062  &  0.47  &   -6.35  &-11.88  \\
47586.8359  &  0.48  &   -6.70  &-11.02  \\
47587.8633  &  0.48  &   -9.53  &-10.34  \\
47598.8789  &  0.53  &  -14.02  & -2.58  \\
47602.8516  &  0.55  &  -15.38  &  0.74  \\
47607.8242  &  0.58  &  -18.86  &  4.20  \\
47608.7930  &  0.58  &  -21.07  &  4.90  \\
47612.8398  &  0.60  &  -21.97  &  8.06  \\
47613.7578  &  0.61  &  -23.18  &  8.32  \\
47628.7344  &  0.68  &  -29.95  & 16.34  \\
47640.7031  &  0.74  &  -32.58  & 18.24  \\
47641.8555  &  0.74  &  -30.13  & 19.15  \\
47642.7695  &  0.75  &  -33.43  & 18.76  \\
47643.7383  &  0.75  &  -32.53  & 18.59  \\
47644.7266  &  0.76  &  -33.16  & 18.78  \\
47661.7305  &  0.84  &  -30.64  & 15.75  \\
47662.6680  &  0.84  &  -30.49  & 15.06  \\
47664.7305  &  0.85  &  -30.27  & 14.10  \\
47665.6602  &  0.86  &  -26.94  & 14.82  \\
47674.5586  &  0.90  &  -24.69  &  8.50  \\
47675.6406  &  0.91  &  -24.16  &  7.88  \\
47676.6641  &  0.91  &  -21.57  &  7.61  \\
47688.5938  &  0.97  &  -14.10  & -2.76  \\
47689.7031  &  0.97  &  -12.11  & -3.75  \\
47693.6289  &  0.99  &  -11.58  & -8.51  \\
47698.6328  &  0.02  &   -7.05  &-12.13  \\
47702.6523  &  0.04  &   -3.51  &-15.23  \\
47723.5664  &  0.14  &   11.06  &-31.07  \\
47730.5547  &  0.17  &   14.62  &-33.99  \\
47763.5078  &  0.33  &   12.28  &-29.11  \\
47879.9688  &  0.89  &  -24.54  &  9.80  \\
47894.9492  &  0.97  &  -14.83  & -2.20  \\
47895.9453  &  0.97  &  -12.76  & -3.57  \\
47898.9492  &  0.99  &  -12.30  & -6.20  \\
47900.9531  &  0.99  &  -10.97  & -9.37  \\
47904.9727  &  0.01  &   -6.57  &-12.07  \\
47908.9570  &  0.03  &   -5.09  &-15.27  \\
47910.9453  &  0.04  &   -3.03  &-16.64  \\
47922.9492  &  0.10  &    7.00  &-27.02  \\
47928.8906  &  0.13  &   10.71  &-30.08  \\
47930.9648  &  0.14  &   13.31  &-30.51  \\
47931.9258  &  0.14  &   14.32  &-31.36  \\
47933.9453  &  0.15  &   12.59  &-32.76  \\
47934.9492  &  0.16  &   15.58  &-33.16  \\
47935.8750  &  0.16  &   15.39  &-33.56  \\
47939.8906  &  0.18  &   15.06  &-34.87  \\
47942.8984  &  0.20  &   16.54  &-35.06  \\
47952.8672  &  0.25  &   16.34  &-35.10  \\
47955.8984  &  0.26  &   16.30  &-34.90  \\
47957.8320  &  0.27  &   14.77  &-35.07  \\
47958.9414  &  0.28  &   16.70  &-34.97  \\
47959.8477  &  0.28  &   15.94  &-33.16  \\
47960.8672  &  0.28  &   15.77  &-33.39  \\
47963.8711  &  0.30  &   14.87  &-33.14  \\
47969.7930  &  0.33  &   13.14  &-30.06  \\
47989.7422  &  0.42  &   -1.21  &-17.35  \\
47990.7617  &  0.43  &   -2.36  &-17.36  \\
47991.7852  &  0.43  &   -3.27  &-16.77  \\
47994.7852  &  0.45  &   -3.84  &-14.83  \\

47998.7109  &  0.47  &   -5.63  &-11.45  \\
47999.7344  &  0.47  &   -5.93  &-10.69  \\
48000.7148  &  0.48  &   -6.97  &-10.60  \\
48001.7500  &  0.48  &  -10.20  & -9.33  \\
48021.6758  &  0.58  &  -19.67  &  3.84  \\
48023.7617  &  0.59  &  -23.76  &  5.77  \\
48026.6797  &  0.60  &  -22.28  &  7.47  \\
48027.6562  &  0.61  &  -21.28  &  8.32  \\
48042.6328  &  0.68  &  -29.91  & 16.44  \\
48044.7461  &  0.69  &  -29.98  & 17.49  \\
48050.6172  &  0.72  &  -32.79  & 18.73  \\
48052.5742  &  0.73  &  -33.89  & 18.41  \\
48054.5898  &  0.74  &  -32.81  & 19.02  \\
48055.6211  &  0.74  &  -34.69  & 18.90  \\
48057.6484  &  0.75  &  -34.80  & 18.49  \\
48058.6992  &  0.76  &  -34.14  & 19.76  \\
48059.6445  &  0.76  &  -35.64  & 17.99  \\
48060.5586  &  0.77  &  -32.41  & 19.65  \\
48069.5781  &  0.81  &  -31.88  & 17.72  \\
48078.6055  &  0.85  &  -28.72  & 14.35  \\
48079.6523  &  0.86  &  -26.86  & 15.33  \\
48082.5820  &  0.87  &  -26.60  & 12.51  \\
48084.5625  &  0.88  &  -26.50  & 10.60  \\
48087.5898  &  0.90  &  -25.65  &  8.99  \\
48088.5781  &  0.90  &  -22.51  &  8.12  \\
48100.5391  &  0.96  &  -14.22  & -1.31  \\
48101.5391  &  0.97  &  -14.20  & -3.01  \\
48102.5781  &  0.97  &  -11.95  & -3.36  \\
48104.5508  &  0.98  &  -11.71  & -6.25  \\
48105.5273  &  0.98  &  -10.52  & -6.74  \\
48106.5273  &  0.99  &  -11.12  & -7.38  \\
48108.5312  &  1.00  &   -8.76  & -8.89  \\
48116.5391  &  0.04  &   -0.66  &-15.99  \\
48280.9727  &  0.83  &  -31.12  & 16.27  \\
48281.9727  &  0.84  &  -30.50  & 15.95  \\
48283.9688  &  0.85  &  -30.04  & 15.62  \\
48289.9531  &  0.88  &  -26.94  & 11.90  \\
48290.0078  &  0.88  &  -28.96  & 11.94  \\
48291.8945  &  0.89  &  -26.25  & 11.06  \\

\hline
\enddata
\end{deluxetable}

\begin{deluxetable}{lll}
\footnotesize
\tabletypesize{\scriptsize}

\tablecaption{IOTA Observing Log of \lamvir.
\label{table_obslog}}
\tablehead{
  \colhead{Date\tablenotemark{a}}     & \colhead{Interferometer}
& \colhead{Calibrator Names}\\   \colhead{(UT)} &
\colhead{Configuration\tablenotemark{b}}  }
\startdata
2003 Feb 16,17  & A35 B05 C10  &  HD 126035 (G7 III, 0.78 $\pm$ 0.24 mas\tablenotemark{c})    \\
                &              &   HD 129502 (F2 III, 1.20 $\pm$ 0.22 mas)    \\
2003 Feb 20-23  & A25 B05 C10  &   HD 126035         \\
 2003 Mar 21     & A35 B07 C25 &   HD 126035          \\
 2003 Mar 22     & A35 B07 C10 &   HD 126035          \\
                 &             &   HD158352 (A8 V, 0.44 $\pm$ 0.10 mas) \\
 2003 Mar 23, 24 & A35 B15 C10 &   HD 126035; HD 158352        \\

2003 Jun 12, 14-16 &A35 B15 C10&   HD 126035         \\
 2003 Jun 17     & A35 B15 C10 &   HD 126035         \\

 2004 Mar 16-21  & A35 B15 C10 & HD 126035          \\
 2004 Apr 13     & A35 B15 C10 & HD 129502          \\
 2004 Apr 14     & A35 B15 C10 & HD 126035; HD 129502          \\
 2004 Apr 20     & A35 B15 C10 & HD 126035; HD 158352          \\
 2004 Apr 24,25  & A35 B15 C10 & HD 126035          \\
 2004 May 28     & A35 B15 C10 & HD 126035          \\
 2004 May 30     & A35 B15 C10 & HD 126035; HD 129502; HD 158352          \\
 2004 Jun 01     & A35 B15 C10 & HD 126035; HD 158352          \\
 2004 Jun 02-07  & A35 B15 C10 & HD 126035          \\

 2005 Jun 14-18  & A35 B15 C10 & HD 126035          \\
\hline
\enddata
\tablenotetext{a}
{Scan Mode 1 before 2003 Jun 16: A: fixed, B: $\Delta X=$50.8 $\micron$ , C: $\Delta X=$25.4 \micron; \\
                   ~~~~~Scan Mode 2 after 2003 Jun 16: A: fixed, B: $\Delta X=$25.4 $\micron$ , C: $\Delta X=$-25.4 $\micron$.}
\tablenotetext{b}{Configuration refers to the location of telescopes A, B, C on
the NE, SE and NE arms respectively; see \citet{Traub2003} for more
details.}
\tablenotetext{c}{Uniform disk (UD) diameters of the calibrators are generally estimated using $getCal$, an SED-fitting
routine maintained and distributed by the Michelson Science Center (http://msc.caltech.edu).}
\end{deluxetable}

\begin{deluxetable}{llll}
\footnotesize \tabletypesize{\small} \tablecaption{Orbital
and Binary Parameters of \lamvir.
\label{bestfit} }

\tablehead{
\colhead{Parameter} &
\colhead{Stickland1975\tablenotemark{a}}  & \colhead{Preliminary
fit\tablenotemark{b}} &\colhead{Best Fit\tablenotemark{c} }}
\startdata
$H$-band Flux ratio      &                 &0.5749 $\pm$ 0.0021      & 0.6055 $\pm$ 0.0056      \\
Period (days)    & 206.64 $\pm$ 0.05    &206.7323 $\pm$ 0.0061     & 206.7321 $\pm$  0.0040   \\
$T_0$ (MJD)       & 40253.1 $\pm$ 15.5   & 53070.28 $\pm$ 0.50      & 53070.30 $\pm$ 0.32       \\
Eccentricity      & 0.079 $\pm$ 0.021    &0.0603 $\pm$ 0.0031      & 0.0610 $\pm$ 0.0036       \\
$\omega$ (degrees)   & 273.3 $\pm$ 26.8      &272.10 $\pm$ 0.71        & 272.28 $\pm$ 0.46         \\
$\Omega$ (degrees)   &                      &196.57 $\pm$ 0.16        & 196.40 $\pm$ 0.22         \\ 
$i$ (degrees)        &                      &109.97 $\pm$ 0.15        & 109.86 $\pm$ 0.24          \\
$a$ (mas)         &                      &19.768 $\pm$ 0.072         & 19.759 $\pm$ 0.079       \\
$K_1$ (km/s)       & 29.51 $\pm$ 0.89     &24.78 $\pm$ 0.17         & 24.78 $\pm$ 0.17          \\ 
$K_2$ (km/s)       & 24.85 $\pm$ 0.65     &27.308 $\pm$ 0.067         & 27.308 $\pm$ 0.067       \\
$\Delta RV$ (km/s) &                 & $-$0.69 $\pm$0.13             & $-$0.70 $\pm$ 0.13        \\
$\gamma$ (km/s) & $-$6.40 $\pm$ 0.41     &$-$8.053 $\pm$0.045              & $-$8.053 $\pm$ 0.045      \\
$f$ coefficient\tablenotemark{d}&        &3.47 $\pm$ 0.18          & 3.08 $\pm$  0.14         \\ 
RV $\chi^2$/dof   &                      &   1.02                  & 1.02                     \\ 
$V^2$ $\chi^2$/dof&                      &  1.40                   & 1.03                      \\ 
CP $\chi^2$/dof   &                      &  1.21                   & 0.12                      \\ 
Total $\chi^2$/dof&                      &   1.21                  & 0.89                      \\ 

\enddata
\tablenotetext{a}{Due to measurement uncertainties in Stickland's work the primary and secondary components are reversed, resulting in a value of $\omega$ that
differs from ours by 180$\arcdeg$. This has been corrected in the table.}
\tablenotetext{b}{Preliminary orbit fit using bandwidth smearing corrected $V^2$ model and re-estimated closure phase errors (see \S\ref{finalorb}).}
\tablenotetext{c}{Closure phases are de-weighted in the best-fit to eliminate biases and uncertainties, especially those in the flux ratio. }
\tablenotetext{d}{The introduced bandwidth smearing coefficient (see \S\ref{vis}).}
\end{deluxetable}

\begin{deluxetable}{lcc}\footnotesize
\tabletypesize{\small}
\tablecaption{Physical Parameters of \lamvir.
\label{tab:physics} }
\tablehead{\colhead{Physical Parameter} & \colhead{Primary Component}  &
\colhead{Secondary Component} }
\startdata
Mass (\msun)\tablenotemark{a}                      & 1.897 $\pm$ 0.016     & 1.721 $\pm$ 0.023  \\
$\pi_{orb}$ (mas)\tablenotemark{a}                 & \multicolumn{2}{c}{18.81 $\pm$ 0.10}       \\
$\pi_{\rm HIP}^{\prime}$(mas)\tablenotemark{b}     & \multicolumn{2}{c}{18.55 $\pm$ 0.84}      \\
System distance (pc)\tablenotemark{a}              & \multicolumn{2}{c}{53.16 $\pm$ 0.29}                 \\
Semimajor axis (AU)\tablenotemark{c}               &\multicolumn{2}{c}{1.0504 $\pm$ 0.0071}    \\    
Visible light ratio                                & \multicolumn{2}{c}{0.56 $\pm$ 0.10}       \\
H band flux ratio\tablenotemark{c}                &  \multicolumn{2}{c}{0.6055 $\pm$ 0.0056} \\
$V$ Magnitude (mag)                                & $5.003 \pm 0.070$    & $5.63 \pm 0.12$     \\
Bolometric flux ($10^{-7}$ erg cm$^{-2}$ s$^{-1}$) & $2.366 \pm 0.010$     & $1.428 \pm 0.089$\\
Total bolometric flux ($10^{-7}$ erg cm$^{-2}$ s$^{-1}$)& \multicolumn{2}{c}{$3.794 \pm 0.014$} \\
Luminosity (\lsun)                                 &$20.84 \pm 0.25$       & $12.58 \pm 0.16$ \\
$T_{eff}$ (K)                                      & 8280 $\pm$ 200        & 8280 $\pm$ 200 \\      
$v \sin i$ (\kms)                                     & 36 $\pm$ 1      & 10 $\pm$ 2   \\

\hline\enddata
\tablenotetext{a}{Parameters that are determined directly from the best-fit orbital parameters.}
\tablenotetext{b}{Revised {\it Hipparcos\/} parallax accounting for orbital motion.}
\tablenotetext{c}{From table \ref{bestfit}}
\end{deluxetable}

\clearpage

\begin{figure}[thb]
\begin{center}
{
\includegraphics[angle=90,width=3.2in]{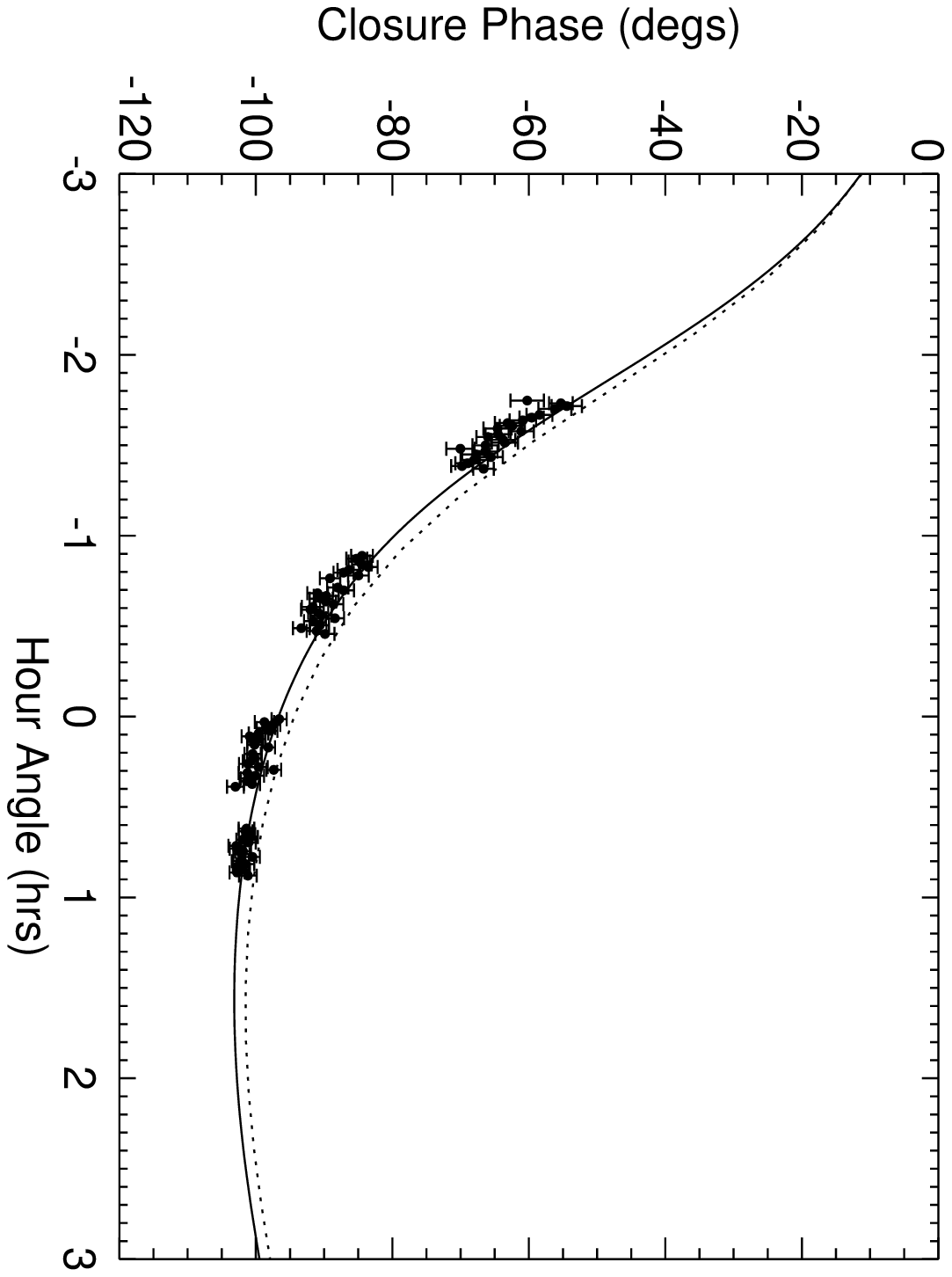}
\includegraphics[angle=90,width=3.2in]{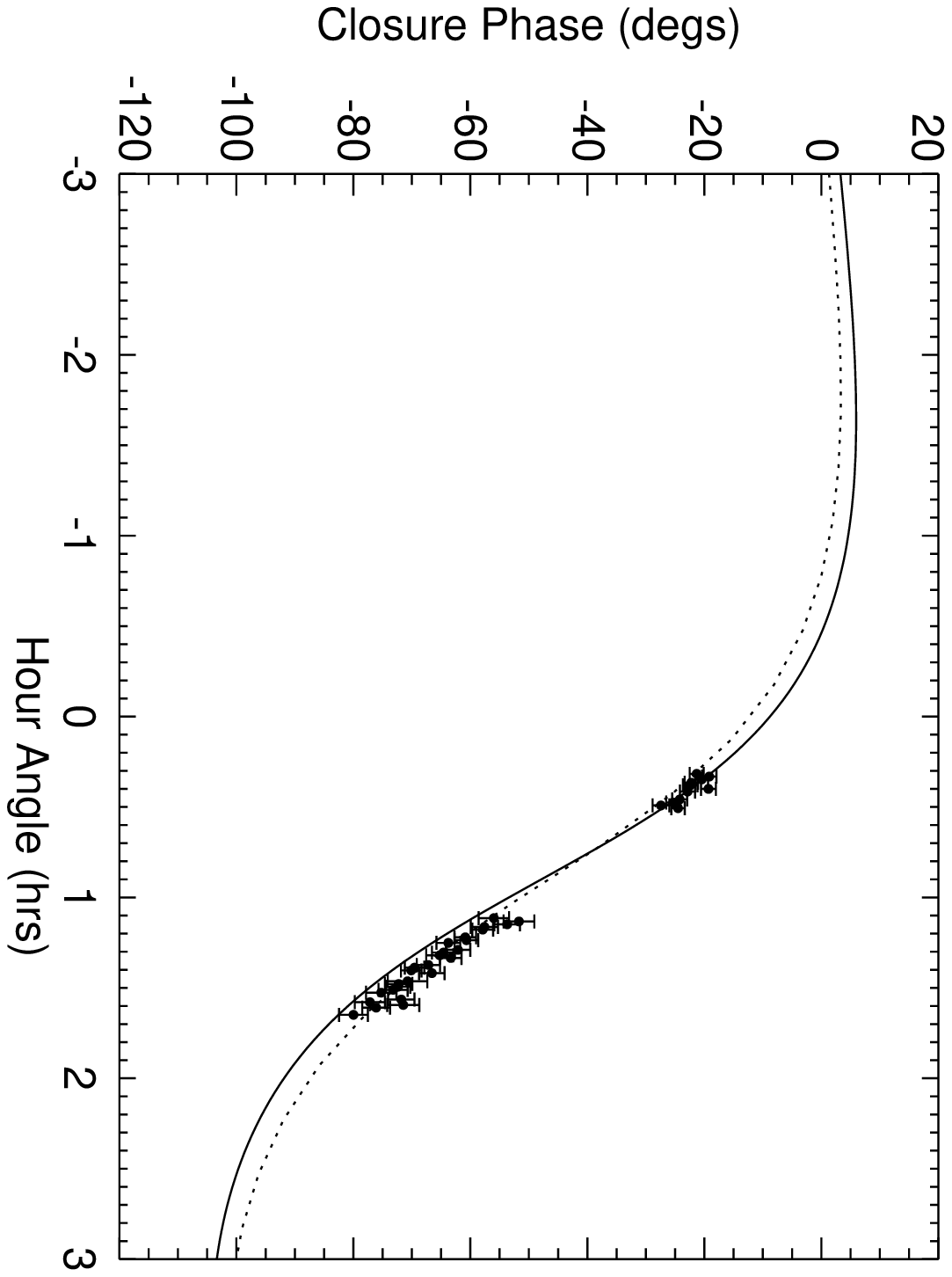}}
\hphantom{.....}
\caption{ 
Preliminary closure phase model vs. hour angle. Two typical dates
of data with different scan modes (left: 2003Mar24, scan mode 1;
right: 2005Jun16, scan mode 2) are selected to represent the whole
data set. The solid lines show the original closure phase model,
while the dotted lines show the model with bandwidth smearing
taken into account. The difference between the two models is about
5 degrees in both panels. Closure phases data are indicated as
filled dots with 1-$\sigma$ measurement errors. \label{smearing1}}
\end{center}
\end{figure}

\begin{figure}[thb]
\begin{center}
{
\includegraphics[angle=90,width=3.2in]{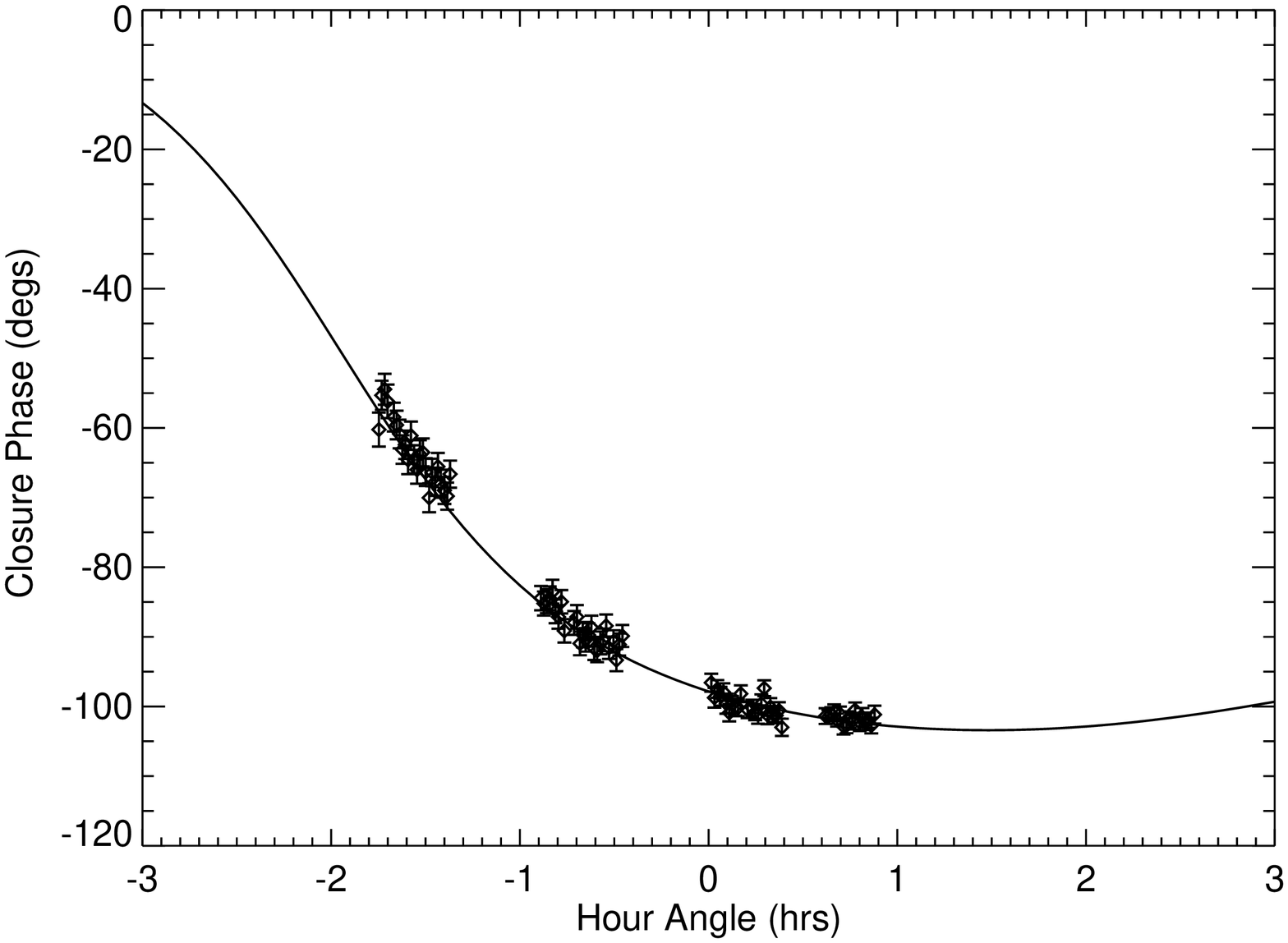}
\includegraphics[angle=90,width=3.2in]{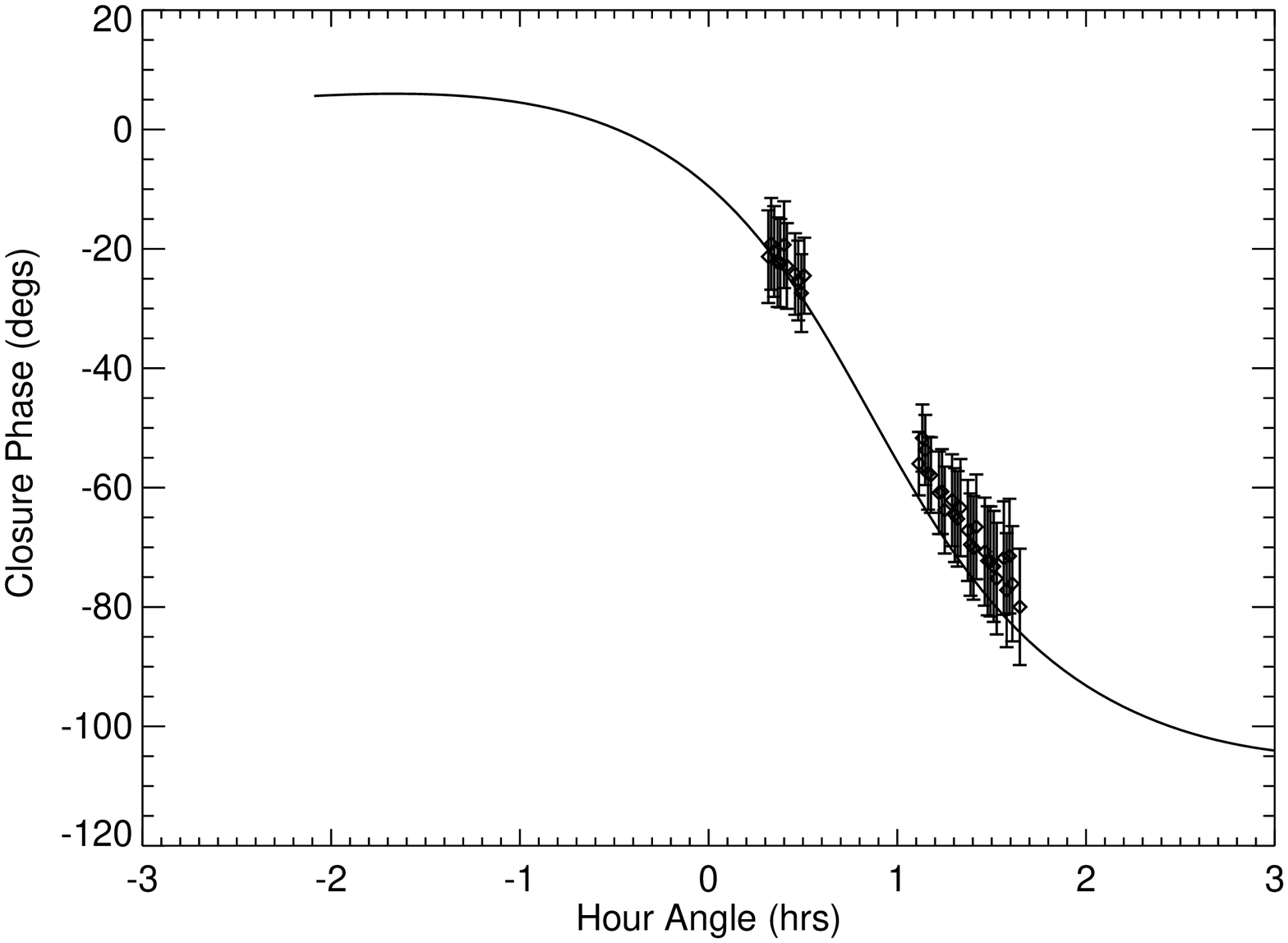}}
\caption{
Closure phase model and data with new estimated errors.
The two panels indicate the same dates as in Fig. \ref{smearing1}.
 The new 1-$\sigma$ errors in the left panel are smaller than those
in the right one due to smaller closure phase fluctuations in
 scan mode 1. The good fit of the data within the
errors suggests the robustness of our error estimation.
\label{smearing3}}
\end{center}
\end{figure}

\clearpage

\clearpage

\clearpage

\appendix

\section{OPD fluctuations and closure phase errors}

  Because fringes are obtained by piezo scanning, the measured
  interferograms are thus temporal sequences that span several
  atmospheric coherence times.  In our data reduction pipeline,
  fringe scans are divided into segments of equal time according to
  the atmospheric coherence time \citep{Baldwin1996} to allow best
  signal-to-noise ratio for averaging the closure phase.  More
  specifically, the complex visibility of each short time segment is
  calculated individually for the 3 baselines. The triple product of
  each segment is thus obtained from a complex multiplication of the
  three visibilities.  Lastly, the (complex) triple products from each
  segment are averaged together with those from the other segments to form a single estimate of the
  complex triple product for each scan.  The closure phase is of
  course derived as the phase argument of the final complex triple
  product \citep[see][]{Monnier1999}.

\begin{figure}[thb]
\begin{center}
{
\includegraphics[angle=0,width=5in]{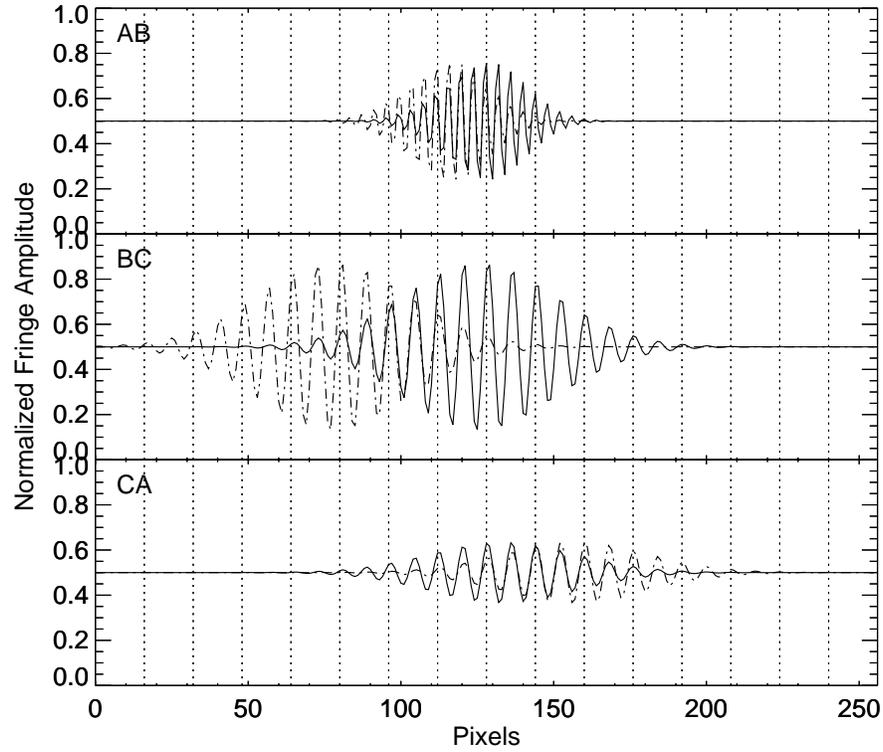}}
\caption{
Simulated \lam~Vir fringes for IOTA baseline AB, BC, 
and CA, respectively. The solid lines indicate normalized fringes with zero 
phase shift while the dotted-dashed lines show an example of fringes 
with phase shifts of 3, $-$6, and 3 wavelengths, respectively. 
The vertical dotted lines divide the fringes into segments of 16 pixels.
\label{fringe}} 
\end{center}
\end{figure}

  Fig.~\ref{fringe} shows an example of this method. The simulated
  fringes are divided into segments of 16 pixels in the pipeline,
  corresponding to 10--20 milli-seconds in time (depending on the scan
  rate).  With zero atmospheric delays, the fringe envelopes are
  aligned in time and the calculation of the triple product is
  straightforward (and the resulting bias on the closure phase has
  already been discussed in \S3.2).  As the atmospheric piston
  fluctuates and causes OPD fluctuations, the fringe envelopes may not
  coincide exactly in time and thus we must consider this additional
  complication on the closure phase estimator.  Because the ``phase''
  of the fringes within the coherence envelope may  not be constant with
  optical path difference, due to both source structure and also due
  to dispersion in beam combining optics, we intuitively can see that
  OPD fluctuations will corrupt the measurement process.  We note that
  this effect does not exist for the monochromatic (i.e., very narrow bandwidth)
  case and we later (see Eq.\ref{eqA}) derive a more quantitiative bandwidth condition.

\begin{figure}[thb]
\begin{center}
{
\includegraphics[angle=90,width=3.2in]{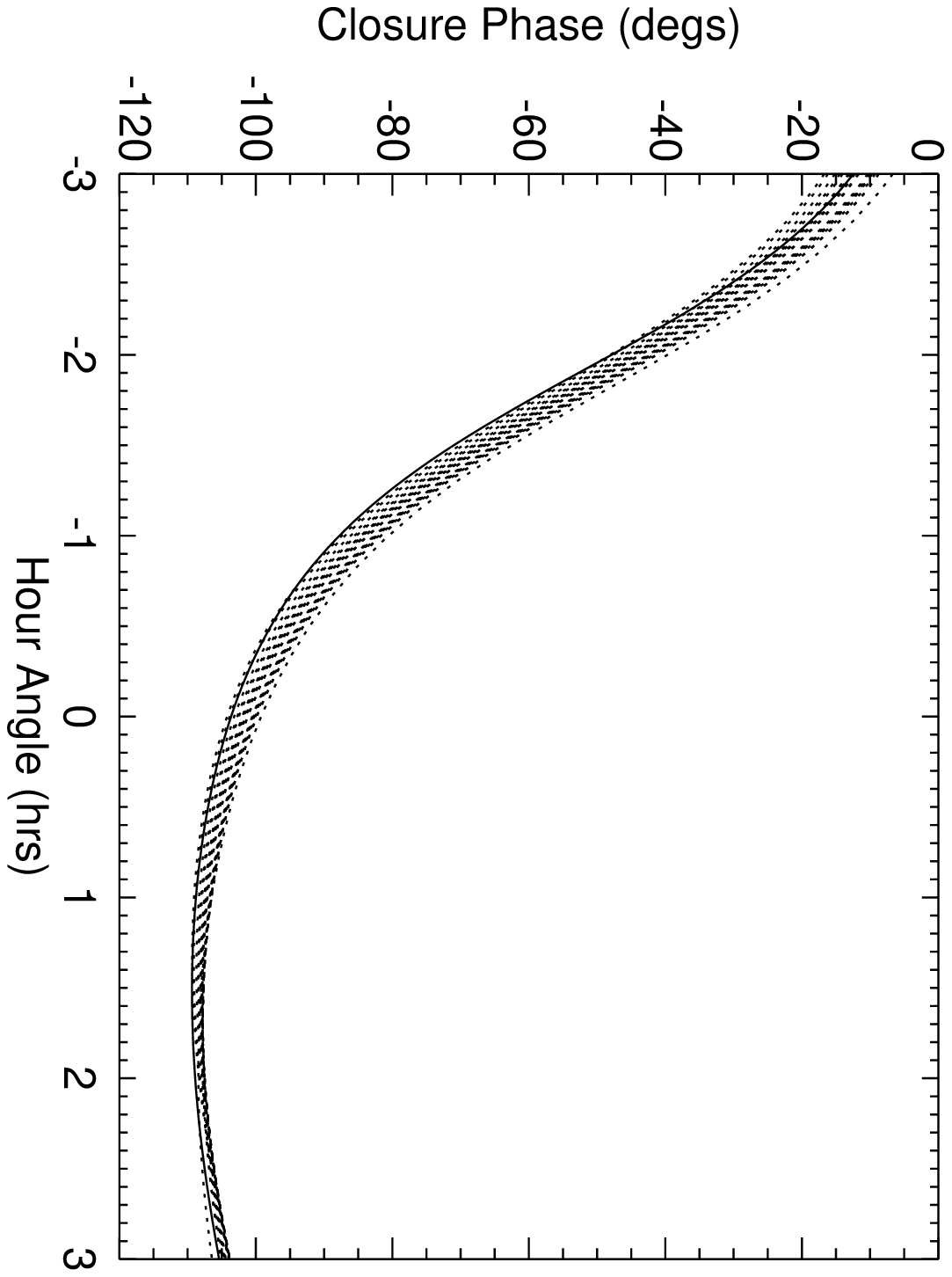}
\includegraphics[angle=90,width=3.2in]{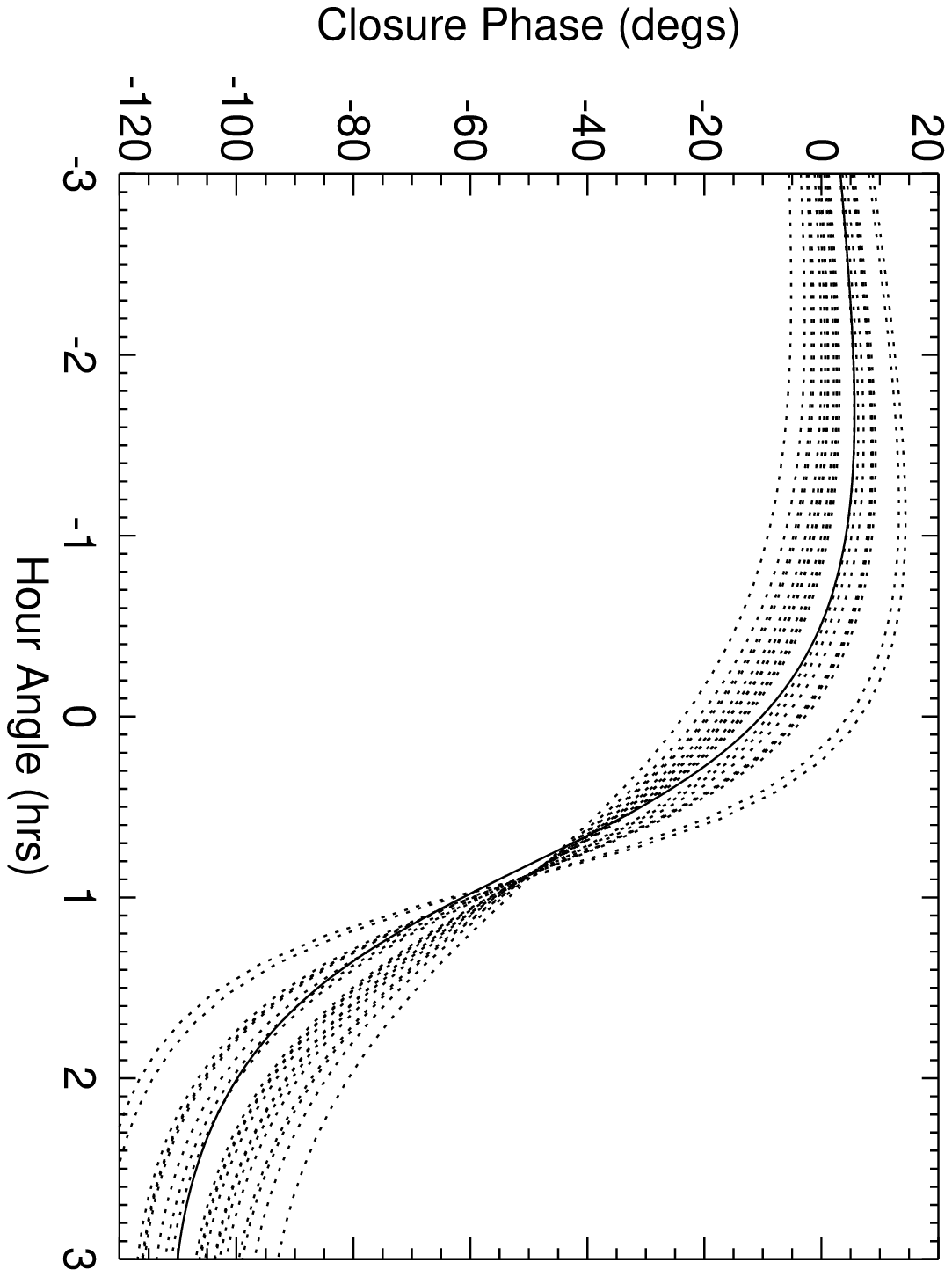}}
\caption{
Closure phase fluctuations due to additional
OPDs caused by the varying atmospheric piston. The
dotted lines indicate simulated closure phases with different
OPDs. The closure phase models with no bandwidth smearing
correction and zero OPD fluctuations are plotted as solid line
for reference. As in Fig.~\ref{smearing1}, two dates with different 
scan modes (left: 2003Mar24, scan mode 1;
right: 2005Jun16, scan mode 2) are selected to represent the entire data.
\label{smearing2}}
\end{center}
\end{figure}

  In order to investigate the influence of this effect, it is
  instructive to consider the case of a binary star and we have
  performed simulations in this Appendix using the parameters of the
  \lamvir\ system.  We have simulated OPD fluctuations above each
  telescope, ensuring the OPDs are closed in triangle (i.e., $OPD_{AB}
  + OPD_{BC} + OPD_{CA}$ = 0). The resulting closure phases are then
  calculated using the IOTA data reduction pipeline (described above
  and also in \S\ref{obs2}).  Fig.~\ref{smearing2} shows 20
  simulated closure phase curves for each of the two representative
  epochs, 2003Mar24 for piezo scan mode 1 and 2005Jun16 for mode 
  2. The closure phases fluctuate in both panels due to the 
  fluctuations of extra OPDs which were assumed to follow a normal
  distribution with 1-$\sigma$ deviation of 1.2 wavelengths
  (determined below).  As can be seen in Fig.~\ref{smearing2}, the
  flucutating OPDs introduce sometimes very large errors in the
  closure phase (up to $\sim10\arcdeg$ in the left panel and
  $\sim30\arcdeg$ in the right) which depend on both hour angle (i.e.,
  projected binary separation) and scan mode (which affects the
  interferograms).  These errors are many
 times larger than those seen from bandwidth smearing calculated without OPD
 fluctuations in Fig.~\ref{smearing1}, suggesting the errors from these
 fluctuations are dominant errors in our closure phase measurements.

 We have reduced the influence of this disturbing effect on our closure phase modeling
 in
 \S\ref{cp} by simulating the closure phase fluctuations for all of our
 data.   From the scatter of simulated closure
 phases, we provide an estimate of the closure phase errors to the model fitting process.
 The standard deviation of the simulated OPD fluctuations was
 determined iteratively in model fitting procedures in order to reduce
 the $\chi_{\nu}^2$ to close to unity.  We found that just a small
 uncorrected atmospheric OPD fluctuation of $\pm$ 2 $\mu m$
 was enough to explain our observed closure phase errors, and this
 level of OPD error is similar to actual residuals reported at IOTA by
 \citet{Pedretti2005}.

\begin{figure}[thb]
\begin{center}
{
\includegraphics[angle=90,width=3.2in]{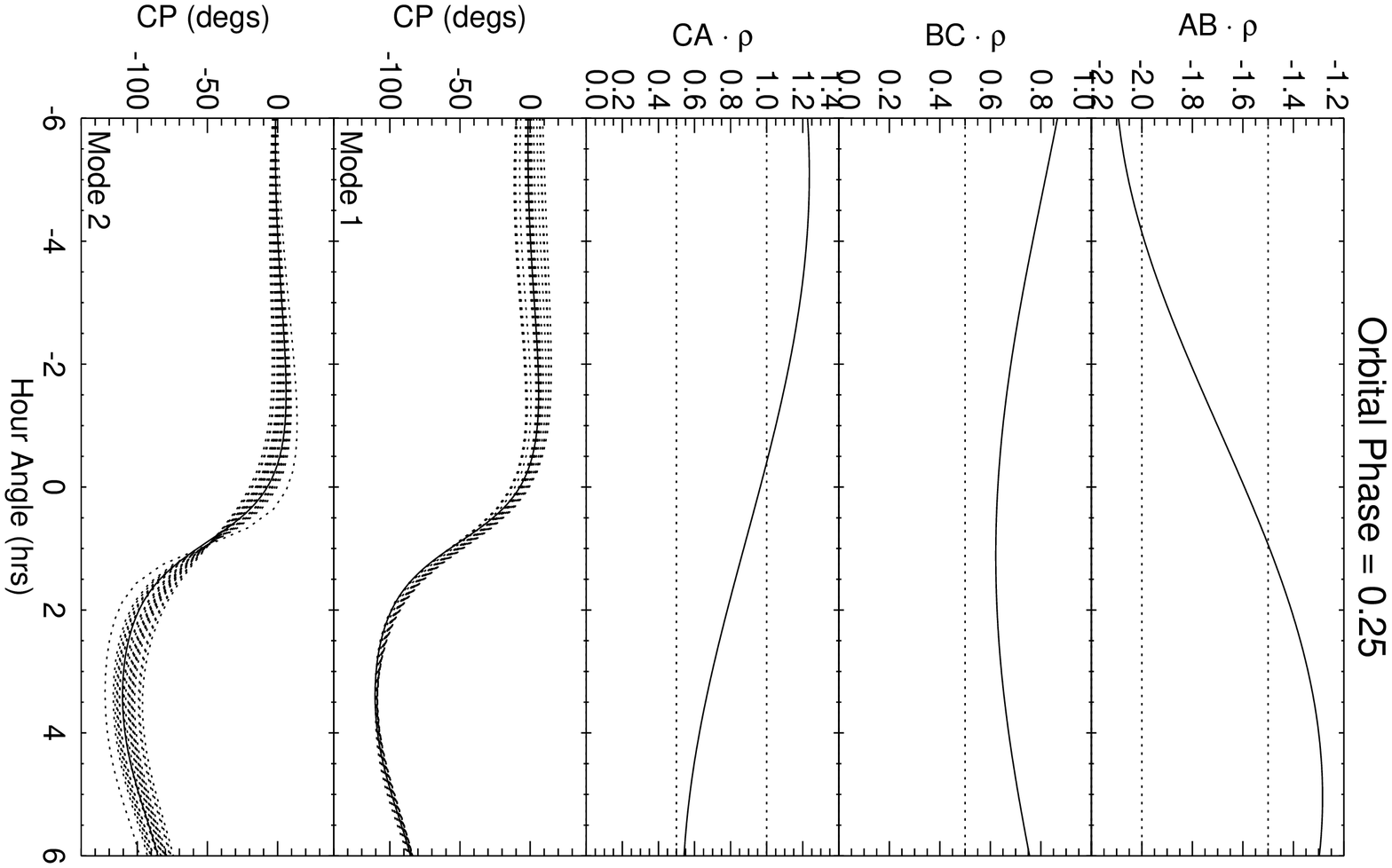}
\includegraphics[angle=90,width=3.2in]{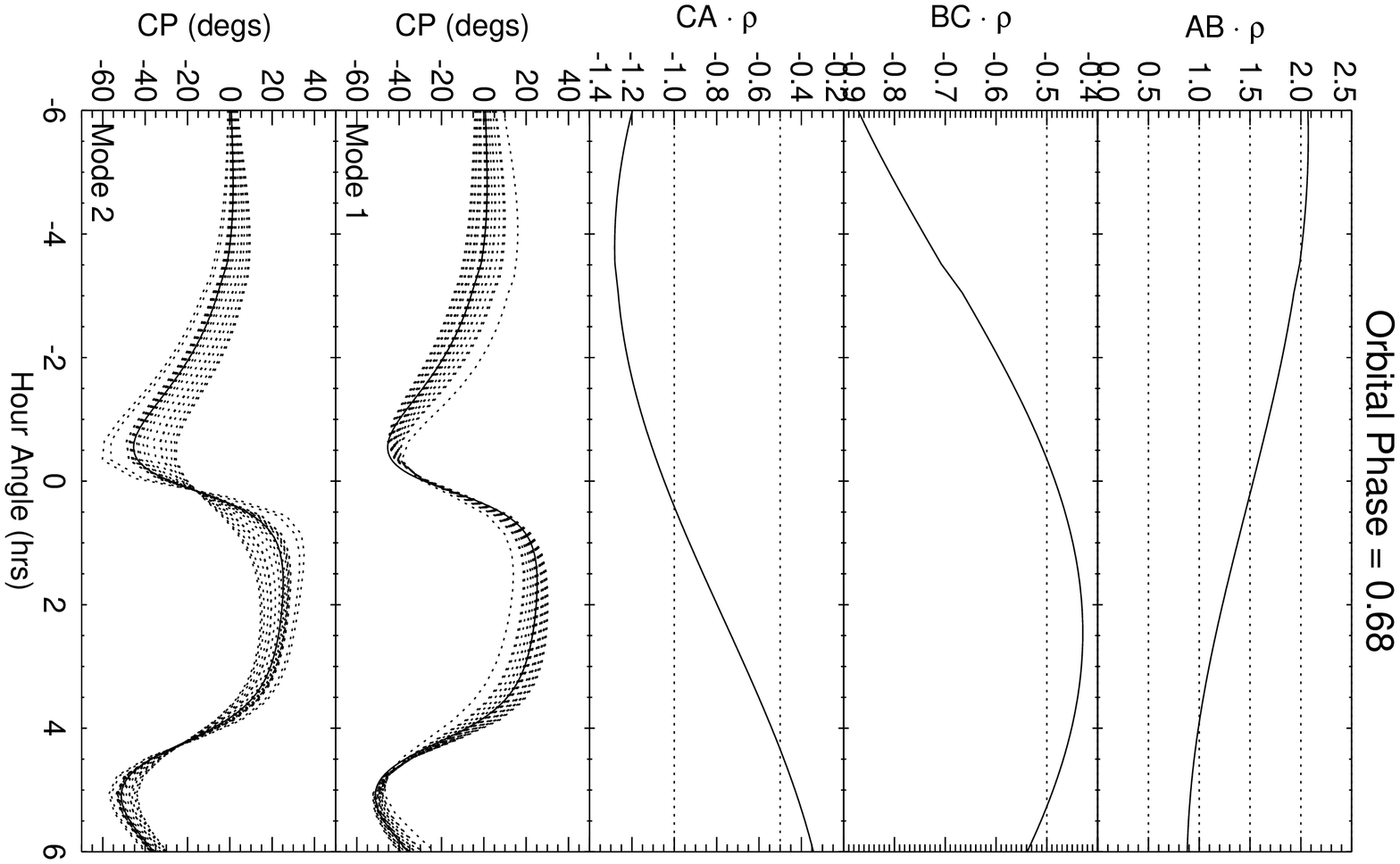}}
\hphantom{.....} 
\caption{
Comparison
of closure phase fluctuations between the two scan modes. We show two typical 
epochs at different orbital phases, the left plot is for phase 0.25, while
the right is for 0.68.  The interferometric 
delay ($\vec{B} \cdot
\vec{\rho}$, in units of wavelength) of the three baselines are plotted as solid lines in the top three panels of each plot. 
The dotted lines indicate phases
of $\pm \frac{n}{2}$ wavelengths, corresponding to $\pm n\pi$ in radian. The
two bottom panels show the corresponding closure phases for scan
mode 1 and 2 respectively. The solid lines indicate the model
closure phase with zero OPD fluctuation and no bandwidth smearing
correction, while the dotted lines indicate the simulated closure phase 
fluctuations. \label{appen1}}
\end{center}
\end{figure}

 Quite unexpectedly, we discovered that the closure phase simulations
 showed ``null points'' where OPD fluctuations had no effect on the
 measured closure phases (e.g., see null fluctuation point in the
 right panel of Fig.\ref{smearing2}, but not in the left panel).  To
 look into this in more detail, we simulated the closure phase
 fluctuations for both scan modes at all observed epochs. Fig.~\ref{appen1} 
 shows two typical epochs and the comparison of the two
 modes at each epoch.  As we can see, the fluctuations of the two
 modes behave differently. Both of them have null points but the
 locations of the nulls are different. This is because fringes are
 scanned from different sides in different modes. For example, in our
 case, one mode scans fringe AC from the left hand side while the
 other scans from the right hand side, which causes the segments of
 fringe AC in the two modes to be scanned at different coherence
 times, thus introducing different errors to the triple products and
 causing the closure phases to fluctuate differently.  For the case of
 a binary star, it is easy to prove that when any two of the 3
 interferometric delays between components ($\vec{B}\cdot\vec{\rho}$)
 are equal to $\pm \frac{n}{2}$ wavelengths, the closure phase will be
 immune to OPD fluctuations and therefore has a null.  The behaviors
 of other nulls may be related to their scan mode and are not yet clear.
 Note that these results are restricted to models of binary
 stars, but presumably apply in general to objects with any resolved
 asymmetric structure affected by bandwidth smearing.

 We can use our empirical study of binary stars to motivate a scaling
 relation for estimating when bandwidth smearing corrupts the closure
 phase measurement process.  In our simulations of IOTA, we found
 strong effects when the source structure (scale: $\rho$) fills
 approximately $\sim\frac{1}{5}$ of the field-of-view defined by
 bandwidth-smearing. Thus, we find that bandwidth-smearing has a much
 stronger and more corruptive effect on closure phases than on
 visibility amplitudes.  We can express this mathematically as a
 condition to meet in order to assure good closure phase measurement:
\begin{equation}
\frac{\vec{B} \cdot \vec{\rho}}{\lambda} < \frac{1}{5}\frac{\lambda}{\Delta \lambda},
\label{eqA}
\end{equation}
where $\vec{B} \cdot \vec{\rho}$  is the previously defined 
interferometric delay; $\vec{B}$ is
the projected baseline vector $(B_x,B_y)$ in unit of meters and
 $\vec{\rho}$ is the binary angular separation ($a$, $b$) in unit of
radian (for cases other than a binary, this represents the typical 
scale of source structure).

In order to avoid these closure phase fluctuations, one could consider
using a closure phase estimator that is not affected by the fringe
phase shift, e.g., an estimator that does not divide fringes into
coherence segements. However, this estimator is likely to be very
noisy unless the entire interferogram is scanned within a coherence
time -- drastically reducing signal-to-noise ratio for faint objects.
Since all of these problems actually stem from bandwidth smearing, we
conclude that using narrow bandwidth is a better approach and is of
importance for precision work.

\clearpage

\clearpage

\end{document}